\documentclass[aps,prx,reprint,twocolumn,floatfix,superscriptaddress,amsmath,amssymb,amsfonts,nofootinbib,balancelastpage]{revtex4-2}

\allowdisplaybreaks

\newcommand{\change}[1]{#1}

\usepackage{color}
\usepackage{url}
\usepackage{graphicx}
\usepackage{amsmath,amssymb,amstext,amsfonts,bm,bbm,dsfont}
\usepackage{enumitem}
\usepackage{hyperref}
\hypersetup{
	colorlinks=true,
	urlcolor=blue,
	citecolor=blue,
	linkcolor=blue,
	breaklinks
}
\usepackage{tabularray}
\usepackage{slashed,physics}
\usepackage{ytableau}
\usepackage{booktabs}
\usepackage[normalem]{ulem}

\usepackage{zref-clever}
\zcsetup{cap,abbrev,nameinlink=false,comp}

\zcRefTypeSetup{equation}{name-sg = equation, name-pl = equations, name-sg-ab = eq., name-pl-ab = eqs.}
\zcRefTypeSetup{figure}{name-sg = figure, name-pl = figures, name-sg-ab = fig., name-pl-ab = figs.}
\zcRefTypeSetup{appendix}{name-sg = appendix, name-pl = appendices, name-sg-ab = app., name-pl-ab = apps.}
\zcRefTypeSetup{section}{name-sg = section, name-pl = sections, name-sg-ab = sec., name-pl-ab = secs.}
\zcRefTypeSetup{table}{name-sg = table, name-pl = tables, name-sg-ab = tab., name-pl-ab = tabs.}

\newcommand{\eqnref}[1]{\zcref{#1}}

\newcommand{\figref}[1]{\zcref{#1}}
\newcommand{\appref}[1]{\zcref{#1}}

\newcommand{\mbf}[1]{\mathbf{#1}}
\newcommand{\vi}{{\boldsymbol{i}}}
\newcommand{\vj}{{\boldsymbol{j}}}

\newcommand{\vk}{{\boldsymbol{k}}}

\renewcommand{\vb}{{\boldsymbol{b}}}

\newcommand{\SO}{\mathrm{SO}}
\newcommand{\SU}{\mathrm{SU}}
\newcommand{\U}{\mathrm{U}}

\newcommand{\vect}[1]{\boldsymbol{#1}}

\begin{document}

\author{Yunchao Zhang}
\affiliation{Department of Physics, Massachusetts Institute of Technology, Cambridge MA 02139-4307, USA}
\affiliation{Department of Physics, Cornell University, Ithaca, NY 14853, USA}
\author{Andreas Feuerpfeil}
\affiliation{Institut für Theoretische Physik und Astrophysik and W\"urzburg-Dresden Cluster of Excellence ctd.qmat, Universit\"at W\"urzburg, 97074 W\"urzburg, Germany}
\author{Subir Sachdev}
\affiliation{Department of Physics and Leinweber Institute for Theoretical Physics, Harvard University, Cambridge MA 02138, USA}
\author{Ronny Thomale}
\affiliation{Institut für Theoretische Physik und Astrophysik and W\"urzburg-Dresden Cluster of Excellence ctd.qmat, Universit\"at W\"urzburg, 97074 W\"urzburg, Germany}
\affiliation{Department of Physics, Indian Institute of Technology Madras, Chennai 600036, India}
\author{Yasir Iqbal}
\affiliation{Department of Physics, Indian Institute of Technology Madras, Chennai 600036, India}

\title{Symmetry-resolved monopoles and stability criteria for an $N_f=12$ Dirac spin liquid on the maple-leaf lattice}

\date{\today}

\begin{abstract}
The $\mathrm{U}(1)$ Dirac spin liquid provides a useful organizing framework for frustrated magnets: it offers an algebraic parent state from which competing orders, confinement patterns, and low-energy spectral features can be understood. Whether such a state can occur as a stable ground state of a two-dimensional spin Hamiltonian remains an open question, because monopole operators of the compact gauge field can proliferate and confine the spinons. Here we determine the symmetry quantum numbers of the fundamental monopoles of a maple-leaf-lattice Dirac spin liquid whose generic six-node regime realizes QED$_3$ with $N_f=12$ two-component Dirac fermions. The charge-one monopoles span a $924$-dimensional complex representation. In the spin-singlet, zero-momentum, zero-$C_6$-angular-momentum sector we find five complex charge-one directions, of which three are mapped to their Hermitian conjugates under time reversal. Their real and imaginary Hermitian combinations therefore supply six independent monopole perturbations that are trivial under translations, $C_6$, time reversal, and spin rotation. The Dirac spin liquid phase is thus not protected by symmetry exclusion of charge-one monopoles: its stability depends on whether these allowed perturbations are relevant in the renormalization group sense. Existing large-$N_f$ and numerical estimates of the charge-one monopole scaling dimension lie close to the relevance threshold, making the maple-leaf lattice a concrete platform for testing the stability of compact QED$_3$ in a quantum magnet. Our complete monopole classification also identifies the spin, momentum, and rotation sectors in which monopole excitations should appear, providing symmetry-resolved benchmarks for exact-diagonalization and variational tests of the $N_f=12$ Dirac spin liquid.
\end{abstract}

\maketitle

\section{Introduction}
\label{sec:intro}
Quantum magnets provide a setting in which the low-energy degrees of freedom need not resemble the microscopic spins from which they arise. In one dimension, this is already familiar from antiferromagnetic spin chains, where the elementary spin-$1$ excitation fractionalizes into spin-$1/2$ spinons~\cite{Faddeev-1981}. In two dimensions, however, deconfined spinons are far more delicate: they require an emergent gauge structure whose fluctuations may either remain deconfined or confine the fractionalized excitations. 
Quantum spin liquids~\cite{Anderson-1973,Balents-2010,Savary-2017,Zhou-2017,Knolle-2019,Broholm-2020} realize the former possibility---they support fractionalized spinons and emergent gauge fields without breaking spin or lattice symmetries. When gapless excitations appear in such states, they must be protected by a mechanism beyond the Goldstone theorem of spontaneous symmetry breaking. In the projective construction, this role is played perturbatively by the projective symmetry group (PSG), which can forbid fermion and gauge-field mass terms and thereby stabilize gapless modes~\cite{Wen-book}. The remaining question is whether these perturbatively stable states also survive the nonperturbative effects of compact gauge fields---a
question that is especially sharp in two spatial dimensions, where monopole events of a compact $\U(1)$ gauge field can destroy the emergent conservation of gauge flux and confine the spinons. 

The $\U(1)$ Dirac spin liquid (DSL) is a canonical example of this
problem~\cite{Hermele-2004,Ran-2007,Hermele2008,Senthil-2005,Senthil-2006,Hermele-2005}.
It is neither a conventional symmetry-breaking state nor a gapped topologically ordered phase, but rather an algebraic spin liquid whose long-range physics 
is described by massless Dirac spinons coupled to an emergent compact $\U(1)$ gauge field.  In continuum notation this theory is QED$_3$,
\begin{equation}
\label{eq:DSL}
    \mathcal{L}_{\rm DSL}
    =
    \sum_{i=1}^{N_f}
    \overline{\psi}_i i\slashed{D}_a \psi_i
    +
    \frac{1}{4e^2} f_{\mu\nu}f^{\mu\nu}\,,
\end{equation}
where \(a_\mu\) is the emergent gauge field,
\(f_{\mu\nu}=\partial_\mu a_\nu-\partial_\nu a_\mu\), and \(N_f\)
counts the number of two-component Dirac fermion flavors.  For
sufficiently many fermion flavors, QED$_3$ is expected to flow in the
infrared to an interacting conformal field theory~\cite{Appelquist-1985,Appelquist-1986,Appelquist-1988,Hermele-2004,Dyer_2013,Karthik-2016a,Karthik-2016b,Chester-2016,Li-2022a,Albayrak-2022,He-2022,Li-2022b}.
When stable, this fixed point describes an algebraic spin liquid with
fractionalized spin-$1/2$ excitations, an emergent photon, and an
infrared symmetry larger than that of the microscopic spin Hamiltonian.

\begin{figure*}
    \centering
    \includegraphics[width=1.0\linewidth]{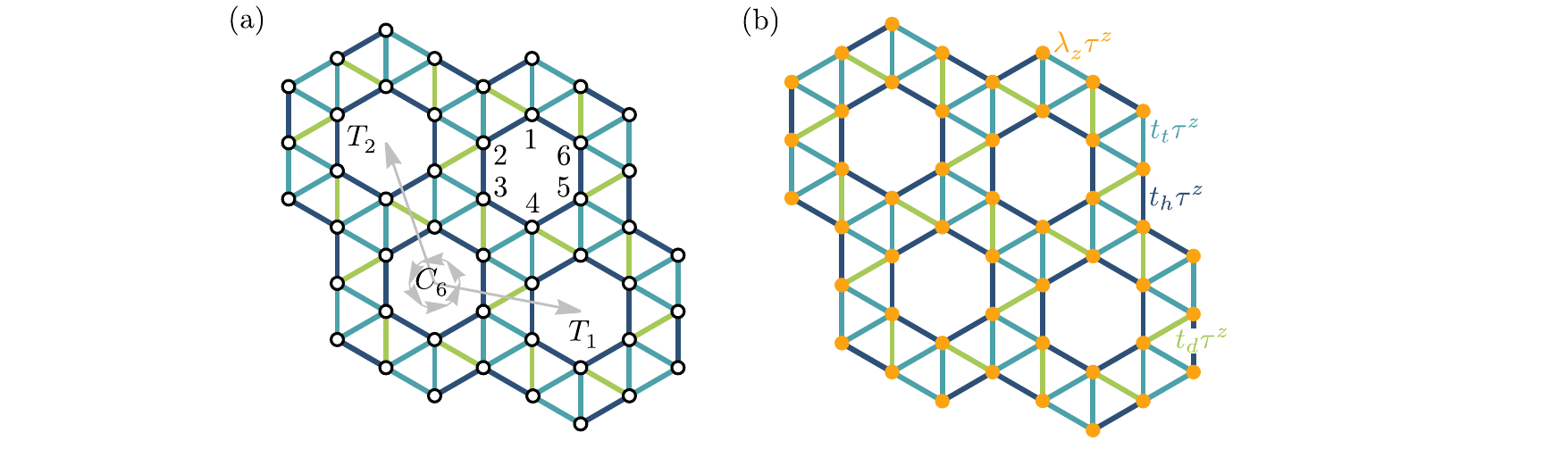}
    \caption{
(a) Maple-leaf lattice and the generators of the microscopic space group used in the symmetry analysis. 
The unit cell contains six sites, and the relevant lattice symmetries are generated by translations
$T_1,T_2$ and a sixfold rotation $C_6$. 
(b) Pattern of mean-field amplitudes defining the uniform $\mathrm{U}(1)$ Dirac spin-liquid ansatz.
The three inequivalent bond amplitudes are denoted by $t_h$, $t_d$, and $t_t$.
}
    \label{fig:u1_qsl_maple}
\end{figure*}

Beyond its role as a candidate infrared phase, the $\U(1)$ DSL provides
a useful organizing framework for frustrated magnets. Its local
operators give a common language for apparently distinct phenomena:
fermion bilinears describe competing conventional orders, monopoles
encode nonperturbative confinement channels, and finite-size spectra
can be compared with the operator content of QED$_3$.

The parton construction also provides a concrete microscopic trial
state. One takes the half-filled Slater-determinant ground state of the
spinon mean-field Hamiltonian and Gutzwiller projects it into the
physical Hilbert space with exactly one fermion per site. This
projected parton state is the variational wave function associated
with the DSL ansatz. It is conceptually distinct from the continuum
QED$_3$ theory, which is the candidate universal long-distance
description of the gapless spinons and emergent gauge fluctuations.
The monopole and bilinear operators of QED$_3$ then organize the
long-distance correlations, competing orders, and confinement
channels expected if a microscopic Hamiltonian is governed by this
infrared theory.

The main obstacle to stability is compactness. In a microscopic spin
system, the emergent gauge flux is not exactly conserved, so the
continuum theory may be perturbed by monopole operators, which insert
\(2\pi q\) units of gauge flux at a spacetime point.  If a
symmetry-allowed monopole is relevant, the emergent
\(\U(1)_{\rm top}\) flux conservation is lost, gauge charges confine,
and the system flows away from the DSL into a conventional ordered state, a symmetric confined phase, or a first-order transition.
Conversely, if all symmetry-allowed monopoles are irrelevant, the compact gauge theory can support a stable algebraic spin liquid.

The role of such tunneling events in quantum antiferromagnets has a long history. In bosonic large-$N$ descriptions, Read and Sachdev showed how Berry phases of hedgehog instantons determine the quantum numbers of monopole events and select the resulting spin-Peierls or valence-bond-solid order~\cite{Read-1990}, while Murthy and Sachdev computed the large-$N$ instanton action in the $(2+1)$-dimensional $\mathrm{CP}^{N-1}$ model~\cite{Murthy-1990}. In fermionic Dirac spin liquids, the corresponding question is whether monopole operators of compact QED$_3$ are allowed by microscopic symmetries and, if so, whether their scaling dimensions exceed the spacetime dimension.

The maple-leaf lattice provides a natural setting in which to revisit this question. It is a $1/7$-depleted triangular lattice with a six-site unit cell and three symmetry-inequivalent nearest-neighbor bonds, conventionally associated with hexagon, triangle, and dimer links. Recent numerical work on spin-$1/2$ Heisenberg models on this lattice has reported a rich competition between canted magnetic order, dimerized paramagnets, and possible
intermediate nonmagnetic regimes~\cite{Ghosh-2022,Gresista-2023,Beck-2024,Nyckees-2025}. This makes the maple-leaf lattice an attractive platform for asking whether a Dirac spin liquid can be stabilized in a microscopic two-dimensional quantum magnet. The specific U(1) Dirac spin liquid studied below has six symmetry-related Dirac valleys; including the physical spin degeneracy, its continuum limit is QED$_3$ with $N_f=12$ two-component Dirac fermions. This large flavor number is the central feature of the present problem: it raises the possibility that even symmetry-allowed fundamental monopoles may be dynamically suppressed.

At the conformal fixed point, the scaling dimension of
the fundamental charge-one monopole has the large-$N_f$
expansion~\cite{Dyer_2013,Borokhov-2002}
\begin{equation}
\Delta_1 = 0.265N_f - 0.0383 + O(1/N_f).
\end{equation}
At $N_f=12$, the known leading expression gives
$\Delta_1=3.1417+O(1/N_f)$. Recent Monte Carlo estimates give
$\Delta_1=2.81(66)$ or $\Delta_1=2.91(41)$~\cite{Karthik-2024}, while an
earlier estimate gives $\Delta_1=3.24(24)$~\cite{Karthik-2019}. These
values overlap the relevance threshold $\Delta_1=3$ within their quoted
uncertainties. Present calculations therefore do not determine the sign
of $\Delta_1-3$ and are compatible with both weak relevance and weak
irrelevance.

In this sense, our analysis follows the same general logic as the early instanton Berry-phase analysis of quantum antiferromagnets~\cite{Read-1990}, but applies it to the monopole operators of a fermionic QED$_3$ Dirac spin liquid. Here, monopole quantum numbers are fixed by fermion zero-modes and by Berry phases of the filled Dirac sea, which can be related to the crystalline topology of the spinon
bands. They determine which monopoles are allowed by microscopic symmetries, which low-energy sectors host the strongest DSL fluctuations, and which confined phases
may appear when the DSL is destabilized~\cite{Song-2019,Song-2020,Zhang-2025,Zhang-2026a}. These quantum numbers also provide sharp diagnostics for numerics:
recent work on the triangular-lattice $J_1$--$J_2$ antiferromagnet
matched Gutzwiller-projected parton states representing the QED$_3$
vacuum and monopole sectors to exact eigenstates in definite momentum,
point-group, and spin sectors~\cite{Wietek24}. A DSL, or a nearby QED$_3$-controlled regime, can thus be identified through the symmetry-resolved structure of the low-energy spectrum~\cite{Budaraju-2023,Wietek24,Budaraju-2025}.

Dirac spin liquids have been proposed in a range of lattice spin
models~\cite{Ran-2007,Iqbal-2013,Iqbal-2014,Iqbal-2016,He-2017,Hu-2019}
and discussed in connection with several experimental
platforms~\cite{Wen-2019,Ding-2019,Bordelon-2019,Zeng-2022,Zeng-2024,Ma-2024,Bag-2024}.
The most studied examples, including the triangular and kagome DSLs,
realize \(N_f=4\)~\cite{Hermele-2005,Ran-2007,Hermele2008,Song-2019,Feuerpfeil_2026,Maity_2026,feuerpfeil_2026_higgs}.
In these cases, stability is helped by the fact that fundamental
monopoles transform nontrivially under microscopic symmetries, so that
the first symmetry-allowed monopole perturbations carry higher monopole
charge and correspondingly larger scaling dimensions.

The maple-leaf lattice~\cite{Betts-1995} [see Fig.~\ref{fig:u1_qsl_maple}] presents the complementary situation: no symmetry excludes the fundamental monopoles, but the flavor number is three times larger. The DSL studied
here arises from a particular fermionic parton ansatz within the broader
PSG classification of maple-leaf spin liquids~\cite{feuerpfeil_2026_higgs,Sonnenschein-2024} and realizes
QED$_3$ with \(N_f=12\), coming from six symmetry-related Dirac valleys
and spin degeneracy. We do not assume that this ansatz is energetically
selected by the nearest-neighbor Heisenberg model; microscopic selection
among competing ansatzes is a separate question. With such a large
flavor number, even symmetry-allowed charge-one monopoles may be
irrelevant, so the stability question is shifted from symmetry exclusion
to the dynamics of a large-flavor QED$_3$ fixed point. This possibility
is timely in view of numerical calculations reporting a putative
nonmagnetic regime in the nearest-neighbor spin-$1/2$ Heisenberg
antiferromagnet on the maple-leaf lattice~\cite{Gresista-2023,Schmoll2025,Ebert2026,Gresista2026,Ebert-2026a}. Our aim is therefore to determine the symmetry-resolved operator content that would have to be observed if a microscopic maple-leaf Hamiltonian realizes this $N_f=12$ regime.

The main result of this work is the microscopic symmetry classification of the charge-one monopoles of the six-node maple-leaf state. The stability-relevant spin-singlet sector at zero momentum and zero $C_6$ angular momentum has complex dimension five. As explained below, microscopic time reversal maps charge-one monopoles to charge-minus-one antimonopoles. After Hermitian conjugation, the corresponding time reversal transformation map has three positive and two negative complex eigen-directions in this five-dimensional sector. For each positive direction $\Phi_{{\rm triv},a}$, $a=1,2,3$, physical time reversal acts as
\begin{equation}
\mathcal T\Phi_{{\rm triv},a}\mathcal T^{-1}=\Phi_{{\rm triv},a}^{\dagger}.
\end{equation}
Consequently both Hermitian combinations
\begin{equation}
\Phi_{{\rm triv},a}+\Phi_{{\rm triv},a}^{\dagger},
\qquad
i\!\left(\Phi_{{\rm triv},a}-\Phi_{{\rm triv},a}^{\dagger}\right)
\end{equation}
are time-reversal even. The maple-leaf DSL therefore admits six independent real Hermitian, fully symmetric charge-one monopole perturbations. Equivalently, the most general such perturbation may be written
\begin{equation}
\delta\mathcal L
=
\sum_{a=1}^{3}
\left(\lambda_a\Phi_{{\rm triv},a}+\lambda_a^*\Phi_{{\rm triv},a}^{\dagger}\right),
\qquad \lambda_a\in\mathbb C.
\label{eq:intro_allowed_monopole_perturbation}
\end{equation}
Thus the noncompact QED$_3$ fixed point is not protected by symmetry exclusion of charge-one monopoles. Its fate is a dynamical question controlled by the scaling dimensions of the allowed perturbations. Existing estimates lie close to marginality and, as discussed above, do not establish whether these operators are relevant or irrelevant at $N_f=12$.

We do not attempt to establish that any particular microscopic
Hamiltonian realizes the maple-leaf DSL.  Our goal is instead to provide
the field-theoretic and symmetry data needed to test such a possibility:
the spin, lattice, and time-reversal quantum numbers of the monopole
operators, together with their implications for stability, low-energy
spectra, and nearby confined phases.  The output of the classification
is therefore a set of spectral predictions.  If a maple-leaf Hamiltonian
enters an $N_f=12$ DSL regime, exact diagonalization should look not
only for the absence of conventional magnetic or valence-bond order, but
also for the symmetry-resolved operator content of QED$_3$: scalar
monopoles in the fully symmetric singlet sector, spinful monopoles in
vector and quadrupolar channels, and fermion bilinears in their
corresponding lattice representations.  Conversely, the absence of this symmetry-resolved organization would place strong constraints on the DSL interpretation of a putative nonmagnetic phase. In this sense, the monopole classification
supplies tests of the maple-leaf DSL that go beyond the variational ansatz itself.

The paper is organized as follows.  In Sec.~\ref{sec:ansatz}, we introduce the fermionic parton construction of the maple-leaf DSL, derive the emergent QED$_3$ description in the infrared, and review the role of monopole operators. In Sec.~\ref{sec:symmetry}, we analyze the microscopic and infrared symmetry structures, determine the embedding of lattice symmetries into the continuum theory, and discuss the associated anomaly constraints. In Sec.~\ref{sec:monopole_quantum_number}, we classify the charge-one monopoles, determine their quantum numbers, and identify the symmetry-allowed monopole perturbations. We conclude in Sec.~\ref{sec:conclusion}.

\section{U(1) Dirac spin liquid on the maple-leaf lattice}
\label{sec:ansatz}

\subsection{\change{Mean-field Hamiltonian}}
We begin by summarizing the fermionic parton ansatz for the maple-leaf
DSL and its continuum limit. To formulate the theory, we employ a parton construction, fractionalizing the spin operators $\vect{S}$ into fermionic spinons $f_{\vi\alpha}, \alpha=\uparrow, \downarrow$ for sites $\vi=(i_x,i_y)$~\cite{Abrikosov-1965}:
\begin{equation}
    \mbf{S}_{\vi}=\frac{1}{2}f_{\vi\alpha}^\dag \bm{\sigma}_{\alpha\beta}f_{\vi\beta}\,.
\end{equation}
Following Wen~\cite{Wen2002}, we introduce a Nambu spinor 
\begin{equation}
    \Psi_{\vi}=\begin{pmatrix}
        f_{\vi\uparrow}\\f_{\vi\downarrow}^\dag
    \end{pmatrix}\,,
\end{equation}
and a mean-field link field $u_{\vi\vj}$, which leads to the Bogoliubov Hamiltonian
\begin{equation}\label{eq:bogoliubov_hamiltonian}
    H_\mathrm{MF}=-\sum_{\vi\vj}\Psi_{\vi}^\dag u_{\vi\vj}\Psi_{\vj}\,.
\end{equation}
Here, 
\begin{equation}
    u_{\vi \vj} = iu_{\vi\vj}^0\tau^0+u_{\vi\vj}^a\tau^a\,,
\end{equation}
with the Pauli matrices $\tau^\mu$ acting on the Nambu spinor $\Psi_{\vi}$. Spin-rotation symmetry implies that $u_{\vi\vj}^\mu \in \mathbb{R}$ and 
\begin{equation}
    u_{\vj\vi}^0=-u_{\vi\vj}^0, \quad  u_{\vj\vi}^a=u_{\vi\vj}^a\,.
\end{equation}
The spinon representation is invariant under $\SU(2)_g$ gauge transformations~\cite{Affleck1988,Dagotto-1988} for a symmetry operator $g$:
\begin{equation}
\begin{split}
    \SU(2)_g: \Psi_{\vi}&\rightarrow W_{g,g(\vi)}\Psi_{g(\vi)}\,,\\
    u_{\vi\vj}&\rightarrow W_{g,g(\vi)}u_{g(\vi),g(\vj)}W^\dag_{g,g(\vj)}\,.
\end{split}    
\end{equation}

\begin{figure*}
    \centering
    \begin{minipage}[b]{0.48\textwidth}
        \includegraphics[width=\textwidth]{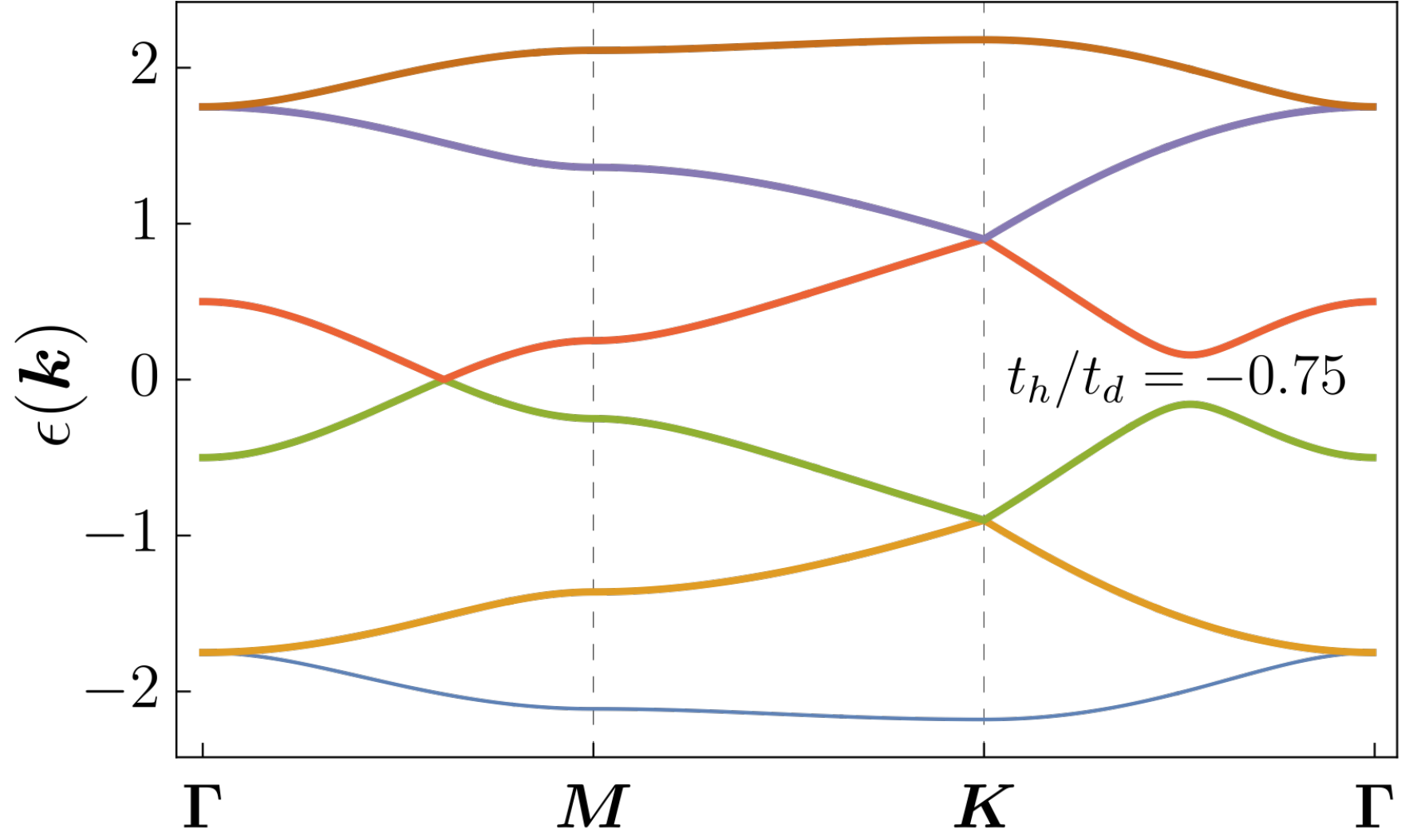}
    \end{minipage}\hfill
    \begin{minipage}[b]{0.48\textwidth}
        \includegraphics[width=\textwidth]{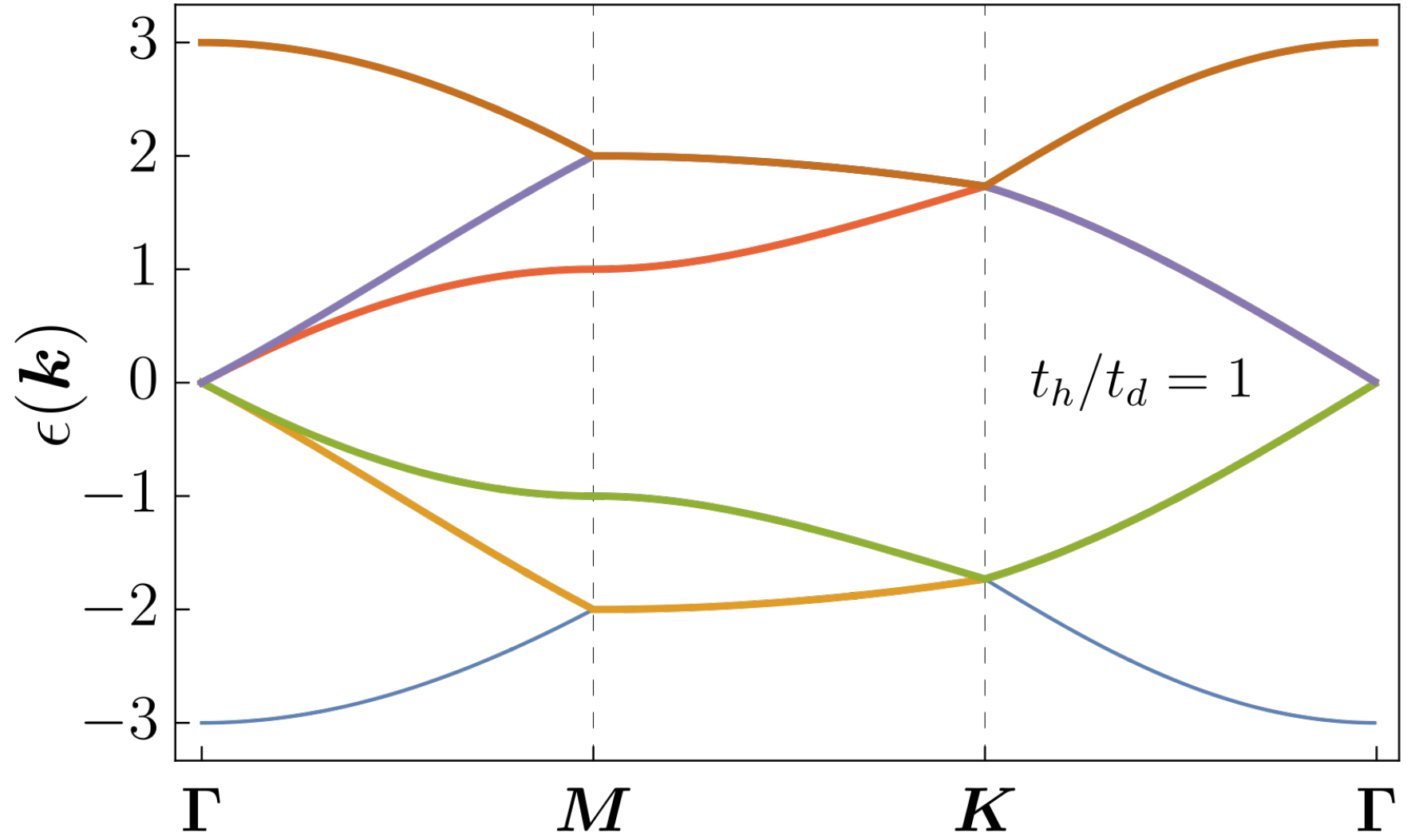}
    \end{minipage}
    \caption{
Mean-field spinon dispersions of the maple-leaf-lattice Dirac
spin liquid obtained from Eq.~\eqref{eq:H_mf}. For
$t_h/t_d=-0.75$ and $t_t=0$, the restricted nearest-neighbor
ansatz contains six Dirac nodes on the
$\overline{\boldsymbol{\Gamma}\mathrm{M}}$ lines, related by $C_6$.
This point lies inside the open six-node region derived in
Eq.~\eqref{eq:six_node_region}. The confinement to these lines is a
property of the nearest-neighbor ansatz and is not enforced by the
microscopic $p6$ space group. This six-valley structure gives the infrared $N_f=12$ QED$_3$ theory studied in the main text.
For comparison, at $t_h/t_d=1$ the spectrum has two Dirac nodes at $\boldsymbol{\Gamma}$;
this distinct $N_f=4$ mean-field DSL is unstable to infinitesimal changes of the hopping parameters.
}
    \label{fig:phase_diag}
\end{figure*}

We adopt a particular $\U(1)$ Dirac spin ansatz~\cite{feuerpfeil_2026_higgs}, whose projective symmetry group (PSG) is outlined in \appref{app:psg}.
The maple-leaf lattice and its associated symmetries are in Fig.~\ref{fig:u1_qsl_maple}. 
\change{Since all link matrices in the present ansatz are pure real hopping, i.e.~$u_{\vi\vj}\propto \tau^z$, we can express the Hamiltonian directly with respect to the spinons $f_{\vi\sigma}$; the Nambu factor then supplies the particle-hole-conjugate block.} Defining the six-component sublattice spinor $f_{\vk\sigma}=(f_{\vk\sigma,1},\ldots,f_{\vk\sigma,6})^{\mathsf T}$ and the phase factors $P=e^{i(3\sqrt{3}k_x-k_y)/2}$ and $Q=e^{i(5k_y-\sqrt{3}k_x)/2}$, the $6 \times 6$ mean-field Hamiltonian reads:
\begin{widetext}
\begin{equation}
\begin{split}
    H_\mathrm{MF}=-\sum_{\vk\change{\sigma}}\change{f^\dag_{\vk\sigma}}
    \begin{pmatrix}
        \lambda_z & t_h & t_t PQ & t_d Q & t_t Q & t_h  \\
        t_h & \lambda_z & t_h & t_t Q & t_d P^{-1} & t_t P^{-1}  \\
        t_t P^{-1}Q^{-1} & t_h & \lambda_z & t_h & t_t P^{-1} & t_d P^{-1}Q^{-1} \\
        t_d Q^{-1} & t_t Q^{-1} & t_h & \lambda_z & t_h & t_t P^{-1}Q^{-1} \\
        t_t Q^{-1} & t_d P & t_t P & t_h & \lambda_z & t_h  \\
        t_h & t_t P & t_d PQ & t_t PQ & t_h & \lambda_z \\
    \end{pmatrix}
    \change{f_{\vk\sigma}}\,.
\end{split}
\label{eq:H_mf}
\end{equation}
\end{widetext}
The Bravais lattice vectors are
$\vect{a}_1=\frac{1}{2}(3\sqrt{3},-1)$ and
$\vect{a}_2=\frac{1}{2}(-\sqrt{3},5)$, while the reciprocal
lattice vectors are
$\vb_1=\frac{2\pi}{21}(5\sqrt{3},3)$ and
$\vb_2=\frac{2\pi}{21}(\sqrt{3},9)$.
Writing $\vk=u\vb_1+w\vb_2$, the phase factors in
Eq.~\eqref{eq:H_mf} become $P=e^{2\pi iu}$ and
$Q=e^{2\pi iw}$.

At the variational level, the mean-field Hamiltonian in Eq.~\eqref{eq:H_mf} defines a half-filled spinon Slater determinant
$\lvert\Psi_{\rm MF}\rangle$. A physical spin wave function is obtained
by enforcing the single-occupancy constraint through Gutzwiller
projection,
\begin{equation}
 \lvert\Psi_{\rm var}\rangle
 =
 \mathcal P_G\lvert\Psi_{\rm MF}\rangle,
 \quad
 \mathcal P_G
 =
 \prod_i\delta_{\hat n_i,1},
 \quad
 \hat n_i=\sum_{\sigma}f^\dagger_{i\sigma}f_{i\sigma}.
 \label{eq:gutzwiller-projected-state}
\end{equation}
It is this projected state, rather than the continuum QED$_3$
Lagrangian, that constitutes a variational wave function for a
microscopic spin Hamiltonian. The QED$_3$ theory derived below is the
candidate infrared description associated with the six-node parton
ansatz after the emergent gauge fluctuations are included.

\subsection{\change{Position of Dirac cones}}
Although the microscopic wallpaper group $p6$ contains no reflection
or glide symmetry and therefore does not generically pin a Dirac node
to a high-symmetry line, the restricted nearest-neighbor Hamiltonian
in Eq.~\eqref{eq:H_mf} has an additional algebraic structure. Within
this three-parameter family, the six nodes lie on the
$C_6$-related $\overline{\vect{\Gamma}\mathrm{M}}$ lines. On the
representative line $w=0$, so that $Q=1$ and
$P=e^{2\pi i u}$, the position of the Dirac node is determined exactly
by
\begin{equation}
\cos(2\pi u_D)=
\frac{
(t_h^2-t_t^2)(t_d^3+2t_h^3)
-t_dt_h^2(t_h^2-3t_t^2)
}{
2t_dt_h^4
}
.
\label{eq:exact_dirac_position}
\end{equation}
The corresponding repeated eigenvalue of the $6\times6$ matrix inside
Eq.~\eqref{eq:H_mf} is most conveniently expressed relative to its
common diagonal term $\lambda_z$. Defining the shifted eigenvalue
$\widetilde{\varepsilon}_D\equiv\varepsilon_D-\lambda_z$, we find
\begin{equation}
    \widetilde{\varepsilon}_D
    \equiv \varepsilon_D-\lambda_z
    =-\frac{t_dt_t}{t_h}\,.
\label{eq:exact_dirac_energy}
\end{equation}

In particular, throughout the open parameter region
\begin{equation}
    -1<\frac{t_h}{t_d}<-\frac{1}{2}\,,
    \qquad
    \left|\frac{t_t}{t_d}\right|
    <
    \left|\frac{t_h}{t_d}\right|\,,
\label{eq:six_node_region}
\end{equation}
Eq.~\eqref{eq:exact_dirac_position} gives
$0<u_D<1/2$ and hence six isolated $C_6$-related Dirac nodes.
The point used in Fig.~\ref{fig:phase_diag},
$(t_h/t_d,t_t/t_d)=(-0.75,0)$, therefore lies in the interior of an
open two-dimensional six-node regime and is not a fine-tuned isolated
point. For $t_t=0$, Eq.~\eqref{eq:exact_dirac_position} reduces to
\begin{equation}
    \cos(2\pi u_D)=
    \frac{t_d^3+2t_h^3-t_dt_h^2}
    {2t_dt_h^2}\,.
\end{equation}
A derivation of Eqs.~\eqref{eq:exact_dirac_position}--%
\eqref{eq:six_node_region} is given in the first subsection of
Appendix~\ref{app:psg}.

We emphasize that the confinement of the nodes to the
$\overline{\vect{\Gamma}\mathrm{M}}$ lines is exact only within the
restricted nearest-neighbor ansatz of Eq.~\eqref{eq:H_mf}; it is not
enforced by the physical $p6$ space group. Generic $p6$-symmetric
longer-range hopping terms may move the nodes away from these lines
without immediately opening a gap.

There is also a separate parameter regime near $t_h=t_d$ that hosts
a different DSL, discussed in Appendix~\ref{app:dsl_2}. That state has
two Dirac nodes at $\vect{\Gamma}$ and is unstable already at the
quadratic mean-field level, because infinitesimal symmetry-preserving
changes of the hopping parameters can gap its spinons.

\subsection{\change{Low-energy theory}}
The low-energy action can be written as a theory of $(2+1)$-dimensional quantum electrodynamics (QED$_3$) with $N_f=12$ flavors of fermions:
\begin{equation}
\begin{split}\label{eq:L_QED}
    \mathcal{L}_\mathrm{QED}&=\sum_{v=1}^{6}\sum_{\sigma=\uparrow,\downarrow}i\overline{\psi}_{v\sigma}\gamma^\mu(\partial_\mu-ia_\mu)\psi_{v\sigma}+\frac{1}{4e^2}f^2+\Delta\mathcal{L}\,,
\end{split}
\end{equation}
where we adopt the Lorentz signature $(1,-1,-1)$ and absorb the Dirac
velocity into the speed of light.  The valley index $v=1,\ldots,6$
labels the six Dirac nodes and $\sigma=\uparrow,\downarrow$ labels the
physical spin, so that the continuum theory contains $N_f=12$
two-component Dirac fermion flavors.  Thus, the six symmetry-related
Dirac valleys of the maple-leaf ansatz provide a microscopic route to
large-flavor QED$_3$, placing the system closer to the regime where
monopoles are expected to be dynamically suppressed.  The gamma matrices
are $\gamma^0=\rho^z$, $\gamma^x=-i\rho^y$, and $\gamma^y=i\rho^x$, with
adjoint $\overline{\psi}=\psi^\dagger\gamma^0$; here the Pauli matrices
$\rho^\mu$ act on the two-component Dirac spinor structure.  As before,
$f_{\mu\nu}=\partial_\mu a_\nu-\partial_\nu a_\mu$ is the field strength
of the emergent $\U(1)$ gauge field $a$.  The correction
$\Delta\mathcal{L}$ contains additional operators---such as velocity
anisotropies---allowed by the microscopic symmetries of the lattice
Hamiltonian.

\subsection{Monopole operators and instabilities of the DSL}
The QED$_3$ action \eqnref{eq:L_QED} admits topologically nontrivial gauge-field configurations~\cite{Borokhov-2002}, described by ``bare'' monopole operators $\mathcal{M}_q$ that insert $2\pi q$ units of $\U(1)$ flux at a spacetime point. $\mathcal{M}^{\dagger}_q$ carry charge $q$ under a global $\U(1)$ symmetry with conserved current equal to the magnetic flux:
\begin{equation}
    \label{eq:gauge_flux}j^{\mu}_{\mathrm{top}}=\frac{1}{2\pi}\epsilon^{\mu\nu\lambda}\partial_{\nu}a_{\lambda}\,.
\end{equation}
We denote this flux symmetry by $\U(1)_{\mathrm{top}}$ to distinguish it from the emergent $\U(1)$ imposed by the parton construction. In the presence of Dirac fermions, the bare monopole operators $\mathcal{M}^{\dagger}_q$ must be dressed by fermion zero-modes (which appear in a nonzero flux background) in order to be gauge invariant. All other local operators, including fermion bilinears $\overline{\psi}_i\psi_j$, can be identified with composites of monopole and antimonopole operators for a given topological charge. We remark that the monopoles are bosonic operators, and QED$_3$ has no local fermions, as fermion parity is part of the $\U(1)$ gauge group.

We focus on the fundamental $(q=1)$ monopoles, which must be
dressed by six of the twelve fermion zero-modes present in the $2\pi$
flux background.  We can therefore represent the gauge-invariant
monopole operators $\Phi$ schematically as
\begin{equation}
    \Phi^{\dagger} \sim d^{\dagger}_i d^{\dagger}_j d^{\dagger}_k
    d^{\dagger}_l d^{\dagger}_m d^{\dagger}_n \mathcal{M}^{\dagger}_1\, .
\end{equation}
The operator $\mathcal{M}^{\dagger}_1$ denotes the bare monopole that
inserts one unit of $2\pi$ gauge flux without filling any zero-mode, while
$d_i^{\dagger}$ creates the fermion zero-mode associated with the parent
Dirac flavor $\psi_i$ in this flux background.  In this notation,
$i,j,\ldots$ label the twelve Dirac flavors of the continuum theory,
equivalently the combined valley and spin labels $(v,\sigma)$; they
should not be confused with the microscopic lattice-site index used
above.  We use a separate symbol $d_i^{\dagger}$, rather than
$\psi_i^{\dagger}$, because the zero-modes in the monopole background
are Lorentz scalars, whereas the parent fields $\psi_i$ are Dirac
spinors.  The zero-modes transform under the flavor symmetry that mixes
the Dirac fermions, but, having vanishing angular momentum in the
unit-flux background, are singlets under Lorentz transformations.

We briefly comment on $\U(1)_{\mathrm{top}}$ neutral operators that are potentially allowed in the QED$_3$ theory if they transform trivially under the microscopic (UV) symmetries. The first class consists of fermion mass bilinears~\cite{Chester-2016b}, which are relevant. As shown in the next section, however, spin rotation, lattice translations, and a momentum-local antiunitary symmetry together forbid all mass bilinears in the interior of the six-node regime. Other classes of operators include four fermion terms and velocity anisotropies, which are already present in the maple-leaf lattice. However, both these have been shown to be irrelevant at the conformal fixed point~\cite{Vafek-2002,Hermele-2005,Zhang-2026b}.

Spinon-pairing perturbations provide yet another class of deformations. They couple the Dirac fermions to charge-two Higgs fields and, when condensed, reduce the gauge structure from $\U(1)$ to $\mathbb{Z}_2$. The resulting Higgs descendants of the maple-leaf DSL were analyzed in Ref.~\cite{feuerpfeil_2026_higgs} and are not our focus here.

\section{Microscopic and infrared symmetries}
\label{sec:symmetry}

We now determine how the microscopic symmetries of the maple-leaf spin
model act on the monopole operators of the infrared (IR) QED$_3$ theory.  We
denote the microscopic (UV) symmetry group by $G_{\rm UV}$.  For the
spin-$1/2$ maple-leaf Hamiltonians considered here, $G_{\rm UV}$ is
generated by the wallpaper group $p6$ (see Appendix~\ref{app:psg}),
time-reversal $\mathcal{T}$, and spin rotations $\SU(2)_s$.  The
infrared theory has a larger symmetry group $G_{\rm IR}$.
Apart from Lorentz symmetry and the discrete spacetime symmetries of
QED$_3$, its internal symmetry is
\begin{equation}
    G_{\rm IR}
    =
    \frac{\SU(12)_f \times \U(1)_{\rm top}}{\mathbb{Z}_{12}}\, .
    \label{eq:IR_global_symmetry}
\end{equation}
Here, $\SU(12)_f$ acts on the twelve Dirac fermion flavors, while
$\U(1)_{\rm top}$ is the topological symmetry associated with
conservation of the emergent gauge flux.  The monopole operators carry
charge under $\U(1)_{\rm top}$.  Because they are dressed by fermion
zero-modes, they also carry flavor quantum numbers, but remain Lorentz
scalars.  We derive this symmetry group in detail in
\appref{app:symmetry}.

The microscopic symmetries are embedded into the infrared symmetry
group through a homomorphism
\begin{equation}
    \varphi: G_{\rm UV} \longrightarrow G_{\rm IR}\, .
\end{equation}
Determining this embedding is the central task of this section.  The
flavor part of $\varphi$ follows from the PSG action on the low-energy
Dirac fermions, while the $\U(1)_{\rm top}$ component is a Berry phase
of the filled spinon sea and must be fixed separately.

From the PSG, the microscopic symmetries act on the low-energy Dirac
fermions as follows. Suppressing the standard action of the spatial
symmetries on the spacetime coordinates, we write
\begin{align}
T_i:\quad
\psi_{v\sigma}
&\longmapsto
e^{i\vect K_v\cdot\vect a_i}\psi_{v\sigma},
\qquad i=1,2,
\label{eq:IR-translation}\\
C_6:\quad
\psi_{v\sigma}
&\longmapsto
U_{C_6,v}\psi_{v+1,\sigma},
\label{eq:C6-continuum}\\
\mathcal T:\quad
\psi_{v\sigma}
&\longmapsto
(i\sigma^y)_{\sigma\sigma'}U_{\mathcal T,v}\psi_{\bar v,\sigma'}.
\label{eq:T-continuum}
\end{align}
Here $\mathcal T$ is antiunitary, so complex coefficients are conjugated
under its action. The label $\vect K_v$ is the crystal momentum of the
$v$th Dirac node, $v+1$ labels the node obtained by applying $C_6$, and
$\bar v$ labels the node at $-\vect K_v$. The Pauli matrix $\sigma^y$
acts on physical spin, whereas the $2\times2$ matrices $U_{C_6,v}$ and
$U_{\mathcal T,v}$ act on the two-component Dirac-spinor index. We use
a common Dirac frame in which these matrices are independent of $v$ and
\begin{equation}
U_{C_6,v}=\mathbbm 1_2,
\qquad
U_{\mathcal T,v}=\rho_x.
\label{eq:chosen_Dirac_frame_matrices}
\end{equation}
The general projected sewing matrices and their transformation under a
change of local Dirac frame are given in Appendix~\ref{app:dirac_frames}.

We next consider $\U(1)_{\rm top}$-neutral quadratic perturbations.
Because physical time reversal on the spinons is the standard Kramers
operation and spin rotation is preserved, it may be combined with a
$\pi$ spin rotation to define $\widetilde{\mathcal T}=K$ on the orbital
Bloch Hamiltonian. The twofold rotation is $C_2=C_6^3$, represented in
the six-sublattice Bloch basis by $P_6^3$. Hence
\begin{equation}
\Theta=C_2\widetilde{\mathcal T}=P_6^3K
\label{eq:momentum_local_antiunitary}
\end{equation}
leaves a generic momentum fixed and obeys $\Theta^2=1$. A local two-band
frame may therefore be chosen in which $\Theta=K$, so that the most
general Hermitian Hamiltonian near an isolated node is
\begin{equation}
h_v(\bm q)=d_0(\bm q)\mathbbm 1+d_1(\bm q)\rho_x+d_3(\bm q)\rho_z,
\label{eq:real_two_band_hamiltonian}
\end{equation}
with real $d_\mu$. The unique two-by-two matrix that anticommutes with
the two kinetic matrices is $\rho_y$, which is odd under $K$ and is
therefore forbidden as a fully symmetric spin-singlet single-cone mass.
Translations additionally forbid bilinears mixing distinct generic
valleys because such terms carry momentum $\vect K_w-\vect K_v$. Thus
symmetry-preserving quadratic perturbations can move the Dirac nodes and
change their velocity tensors, but cannot gap the generic six-node
spectrum infinitesimally. At special parameter values where valleys
collide at the same crystal momentum, translation no longer excludes
intervalley mixing and a symmetry-allowed mass may become possible; an
additional two-valley matrix can, for example, make $\rho_y\tau_y$ real.
This establishes quadratic stability of the six-node spectrum in the
interior of the open region, not the nonperturbative stability of the
compact $\U(1)$ DSL against monopoles.

The $q=1$ monopoles transform in the sixfold antisymmetric tensor
representation of $\SU(12)_f$.  In total, there are $924$ monopoles,
which decompose under
$\SU(12)_f\rightarrow \SU(2)_s\times \SU(6)_v$ as
\begin{equation}
    \mathbf{924}_{12} =(\mathbf{7}_{2},\mathbf{1}_{6})\oplus (\mathbf{5}_2,\mathbf{35}_6)\oplus (\mathbf{3}_2,\mathbf{189}_6)\oplus (\mathbf{1}_2,\mathbf{175}_6)\, .
\end{equation}
In this notation the two entries denote irreducible representations of
$\mathrm{SU}(2)_s$ and $\mathrm{SU}(6)_v$, respectively; the bold number
gives the representation dimension, while the subscript labels the
corresponding group. This branching is useful because the PSG action factors naturally
through the valley and spin symmetries. 
We outline this construction in more detail in Appendix~\ref{app:monopole_su12}.

After separating the two-component Dirac-spinor action, the $\SU(12)_f$ symmetry action on the monopole zero modes reduces to the
induced action on the physical-spin and valley-flavor indices. 
This is
possible because the fermion zero modes in the unit-flux background
have vanishing Lorentz angular momentum and are Lorentz scalars.
Accordingly, the zero-mode operators $d_i$ transform like the
corresponding spin--valley flavor components of the Dirac fermions,
but they carry no two-component Dirac-spinor index.

It remains to determine the $\U(1)_{\rm top}$ Berry phases that
accompany the microscopic symmetries.  In particular,
$\varphi(G_{\rm UV})|_{\U(1)_{\rm top}}$ for the lattice symmetries is
obtained by decomposing the filled spinon bands into a sum of atomic
Wannier insulators~\cite{Song-2019}.  This decomposition determines an
effective charge background $\Gamma^{\mathrm{PSG}}$.  The Berry phase
under a lattice symmetry can then be understood as the
Aharonov-Bohm phase accumulated when a monopole moves in this background. Details are given in
Appendix~\ref{app:berry_phase}, where we also
constrain the monopole Berry phases under the IR discrete symmetries
using analytic arguments similar to those of Ref.~\cite{Song-2020}.

The result is that microscopic time reversal reverses the $\U(1)_{\rm top}$ charge. Its reference phase is fixed below using the spin-$3$, valley-singlet monopole. The Wannier-center analysis fixes the \emph{net physical} $C_6$ eigenvalue of this valley-singlet reference multiplet to $-1$; how this sign is divided between the flavor lift and the $\U(1)_{\rm top}$ factor is convention dependent. In the convention used below, the spin-$3$ reference monopole is mapped by time reversal to minus its Hermitian conjugate. The Wannier-center decomposition gives trivial Berry phases for translations.\footnote{The translation Berry phases can also be deduced from the space-group relation
\begin{equation}
    C_6T_1C_6^{-1}=T_1T_2\,.
\end{equation}
Acting on the valley-singlet monopoles that transform under
$(\mathbf{7}_{2},\mathbf{1}_{6})$, $C_{6},T_{1,2}$ act only as Berry
phases $e^{i\theta_{C_6,T_{1,2}}}$. The above relation immediately
constrains $\theta_{T_2}=0$. Similarly,
\begin{equation}
    T_1 C_6 T_2 = C_6\implies \theta_{T_1}=0\,.
\end{equation}}

To fix the convention used in the monopole classification, recall the
action of $C_6$ on the six valley labels in Eq.~\eqref{eq:C6-continuum},
which is the permutation
$\mathcal P_{C_6}=(1,2,3,4,5,6)$. Although
$\det\mathcal P_{C_6}=-1$, the full spin--valley matrix
$\mathbbm 1_2\otimes\mathcal P_{C_6}$ has unit determinant and is an
element of $\SU(12)_f$. The net physical $C_6$ eigenvalue of the spin-$3$, valley-singlet
reference multiplet is $-1$. With the full $\SU(12)_f$ lift
$\mathbbm 1_2\otimes\mathcal P_{C_6}$ used here, its exterior-power
action already supplies this sign, so the accompanying
$\U(1)_{\rm top}$ factor is $B_{C_6}=1$. The net physical rotation
acting on charge-one monopoles is therefore
\begin{align}
\widehat C_6^{(q=1)}
&=B_{C_6}\,\wedge^6\!\left(
\mathbbm 1_2\otimes\mathcal P_{C_6}
\right)\nonumber\\
&=\wedge^6\!\left(
\mathbbm 1_2\otimes\mathcal P_{C_6}
\right).
\label{eq:net_C6_monopole}
\end{align}
All angular-momentum labels $\eta\in\mathbb Z_6$ below are defined by
the eigenvalues $e^{i\pi\eta/3}$ of the net operator in
Eq.~\eqref{eq:net_C6_monopole}. A detailed construction of the
time-reversal action on monopoles is given in
Appendix~\ref{app:monopole_representation}.

\subsection{Anomaly constraints on the IR}

As a nontrivial check of our results, we now use 't Hooft anomaly
matching to show that the nontrivial $C_6$ embedding---equivalently, the
net $-1$ action on the spin-$3$, valley-singlet reference monopole---arises
naturally from matching the UV lattice system to the IR QED$_3$ theory.  With six
sites per maple-leaf unit cell, each translation unit cell carries an
integer spin.  Therefore, there are no
Lieb-Schultz-Mattis-Oshikawa-Hastings
(LSMOH)~\cite{Lieb-1961,Oshikawa-2000,Hastings-2004} constraints coming
from the interplay between lattice translations and spin symmetry.  The
same is true for $C_6$ rotation.  Nevertheless, even in the absence of a LSMOH anomaly, the embedding of the microscopic symmetries
into the infrared symmetry group is still constrained.  In particular,
the form of $\varphi(G_{\mathrm{UV}})$ must be such that the anomaly of
the IR theory pulls back to the UV anomaly.

Concretely, the LSMOH anomaly matching condition requires the
IR anomaly
$\Omega_{\mathrm{IR}}\in H^4(G_{\mathrm{IR}},\U(1))$ to pull back to the
UV anomaly
$\Omega_{\mathrm{UV}}\in H^4(G_{\mathrm{UV}},\U(1))$ under
$\varphi$~\cite{Zou-2021,Ye-2022,Liu-2025,Zhang-2026b},
\begin{equation}
    \varphi^*[\Omega_{\mathrm{IR}}]=\Omega_{\mathrm{UV}}\,.
\end{equation}
This anomaly matching condition gives an intrinsic characterization of
the IR theory, independent of any mean-field construction.  Since
$\Omega_{\mathrm{UV}}=0$ in the present case, the pullback of the IR
anomaly must vanish.

We now show how this cancellation occurs.  For convenience, we further
split the valley symmetry
$\SU(6)_v\subset \SU(12)_{f}$ into
$\SU(2)_v\times \SU(3)_v$.  Coupling the theory to background
$\mathrm{PSU}(2)_s$, $\mathrm{PSU}(2)_v$, and
$\U(1)_{\mathrm{top}}$ gauge fields, the relevant part of the bulk
$(3+1)d$ QED$_3$ anomaly action over a closed $4$-manifold $M_4$ can be
written as~\cite{Song-2020,Calvera-2021,Zhang-2026a}
\begin{equation} 
    \frac{S_{\rm bulk}}{2\pi i}=\int_{M_4}\frac{1}{2}w_2^v\cup w_2^s+\frac{1}{2}\frac{\mathrm{d}\mathcal{A}^{\mathrm{top}}}{2\pi}\cup w_2^s+\dots\pmod{1}\,,
\end{equation}
where $\dots$ are emergent anomalies in the IR that pull back to zero in the UV. Here, $w_2^{s/v}\in H^2(M_4,\mathbb{Z}_2)$ are the obstruction classes to lifting the $\mathrm{PSU}(2)=\SO(3)$ gauge fields to $\SU(2)$ gauge fields, while $\mathcal{A}^{\mathrm{top}}$ is the $\U(1)_{\mathrm{top}}$ gauge field.

The two displayed terms have different physical meanings.  The first
term in the QED$_3$ anomaly action is a generalized form of the parity
anomaly.  The second term is a mixed flavor and
$\U(1)_{\mathrm{top}}$ anomaly, constraining the magnetic particles of
$\U(1)_{\mathrm{top}}$---the Dirac fermions $\psi$---to carry spin-$1/2$
under $\SU(2)_s$.

We next evaluate the pullback of this anomaly under the embedding
$\varphi:G_{\mathrm{UV}}\rightarrow G_{\mathrm{IR}}$.  The $C_6$ action
on $\SU(2)_v$ leads to
\begin{equation}
    \varphi^*(w_2^v)=c^2\,.
\end{equation}
The cohomology class $c\in H^1(p6,\mathbb Z_2)$ generates the
$\mathbb Z_2$ gauge field associated with $C_2$ rotation; we write
$c^2\equiv c\cup c$ for its cup square. Substituting this into the first term of the anomaly action gives a contribution
$\frac{1}{2}w_2^sc^2$.  For the pullback of the full IR anomaly to
vanish, this term must be canceled by the Berry phase contribution from
$\U(1)_{\mathrm{top}}$.

To display this cancellation, we recall that $C_6$ acts as $-1$ on the valley singlet monopoles. 
Therefore, in addition to its action in $\SU(12)_f$, it must act with an additional $\U(1)_{\mathrm{top}}$ factor of $-1$.\footnote{This is consistent with Eq.~\eqref{eq:net_C6_monopole}, if instead the
valley permutation is lifted by the center factor
$e^{i\pi/6}\mathbbm1_{6}$, so that the valley action is properly valued in $\SU(6)_v$ and respects the branching imposed by our anomaly analysis.  By the diagonal $\mathbb Z_{12}$
identification in Eq.~\eqref{eq:IR_global_symmetry}, this changes the accompanying
$\U(1)_{\mathrm{top}}$ factor from $B_{C_6}=1$ in
Eq.~\eqref{eq:net_C6_monopole} to $-1$, without changing the physical
$C_6$ action.}
Consequently, the $\U(1)_{\mathrm{top}}$ Berry
phase imposes
\begin{equation}
    \varphi^*(\mathrm{d}\mathcal{A}^{\mathrm{top}})=2\pi c^2\,.
\end{equation}
The second term in the bulk anomaly action therefore contributes the
same class, $\frac{1}{2}c^2\cup w_2^s$, and the two terms cancel
modulo one.  The result is\footnote{To be completely rigorous, what we
have done is not to check
$\varphi^*[\Omega_{\mathrm{IR}}]=\Omega_{\mathrm{UV}}$, but instead
calculating $\varphi^*(L_{\mathrm{IR}})=L_{\mathrm{UV}}$, where
$L\in H^2(G,\mathbb{Z}_2)$ and
$\Omega_{\mathrm{UV/IR}}=\iota(L_{\mathrm{UV/IR}})$ under the inclusion 
\begin{equation}   
    \iota\;:\;H^*(G,\mathbb{Z}_2)\rightarrow H^*(G,\U(1))\,.
\end{equation} 
However, it turns out checking
$\varphi^*(L_{\mathrm{IR}})=L_{\mathrm{UV}}$ is a sufficient (though
not necessary) condition for anomaly matching~\cite{Ye-2022}.}
\begin{equation}
   \frac{ \varphi^*(S_{\mathrm{bulk}})}{2\pi i}=\int_{M_4}\frac{1}{2}c^2\cup w_2^s+\frac{1}{2}c^2\cup w_2^s+\dots=0\pmod{1}\,.
\end{equation}
Thus, the pullback of the IR anomaly vanishes, as required by
$\Omega_{\mathrm{UV}}=0$.  This provides a nontrivial anomaly-matching
check of the $C_6$ Berry phase obtained from the Wannier-center
analysis.

\section{Monopole quantum numbers}
\label{sec:monopole_quantum_number}

\subsection{Complex decomposition and counting conventions}
\label{sec:monopole_dimension_counting}

The charge-one monopoles form the complex vector space
\begin{equation}
\mathcal H_{q=1}
=
\wedge^6\!\left(\mathbb C_s^2\otimes\mathbb C_v^6\right),
\qquad
\dim_{\mathbb C}\mathcal H_{q=1}
=\binom{12}{6}=924.
\label{eq:monopole_complex_space}
\end{equation}
Under $\SU(12)_f\rightarrow\SU(2)_s\times\SU(6)_v$, this space
branches as
\begin{equation}
\begin{split}
\mathbf{924}_{12}={}&
(\mathbf 7_2,\mathbf 1_6)
\oplus(\mathbf 5_2,\mathbf{35}_6)\\
&\oplus(\mathbf 3_2,\mathbf{189}_6)
\oplus(\mathbf 1_2,\mathbf{175}_6).
\end{split}
\label{eq:branching_main}
\end{equation}
Here $\mathbf{2S+1}_2$ is the spin-$S$ irrep of $\SU(2)_s$, while
the second entry is the parent valley irrep of $\SU(6)_v$. The four
complex block dimensions are respectively $7$, $5\times35=175$,
$3\times189=567$, and $175$, which sum to $924$.

Table~\ref{tab:monopole_decomposition} gives a disjoint decomposition
of these complex spaces under the net physical $C_6$ action. For a
fixed spin $S$ and angular momentum $\eta$, $m_\eta$ denotes the full
complex dimension, including all $2S+1$ spin components. The quantity
$m_{\Gamma,\eta}$ is the dimension of its intersection with the joint
translation-trivial subspace $T_1=T_2=1$, and
$m_{\mathrm{non}\text{-}\Gamma,\eta}=m_\eta-m_{\Gamma,\eta}$ is the
remaining complex dimension. Since
\begin{equation}
C_6T_1C_6^{-1}=T_1T_2,
\qquad
C_6T_2C_6^{-1}=T_1^{-1},
\label{eq:p6_translation_relations_table}
\end{equation}
$C_6$ and translations cannot in general be diagonalized
simultaneously at nonzero momentum. Thus
$m_{\mathrm{non}\text{-}\Gamma,\eta}$ is a dimension difference, not
a claim that the corresponding states possess simultaneous
$C_6,T_1,T_2$ eigenvalues.

Microscopic time reversal reverses the topological charge and therefore maps the charge-one monopole space to the charge-minus-one antimonopole space. It should not be interpreted as an antiunitary endomorphism of a fixed $q=+1$ sector. Because the sixfold antisymmetric representation of $\SU(12)_f$ is self-conjugate, the invariant tensor $E$ identifies the conjugate antimonopole representation with the same lower-index $924$-dimensional coefficient space. This defines a \emph{linear} coefficient-space map $\mathsf A_{\mathcal T}$, given explicitly in Appendix~\ref{app:monopole_representation}. We denote by $d_+$ and $d_-$ the complex dimensions of its $+1$ and $-1$ eigenspaces within a specified spin and $C_6$ sector; their intersections with the translation-trivial subspace are denoted $d_{\Gamma,+}$ and $d_{\Gamma,-}$. These are bookkeeping dimensions of coefficient vectors, not dimensions of independent charge-one time-reversal eigenoperators. Each complex $+$ direction produces two real Hermitian time-reversal-even operators, while each complex $-$ direction produces two real Hermitian time-reversal-odd operators. Physical time reversal maps a charge-one monopole with $C_6$ eigenvalue $e^{i\pi\eta/3}$ to an antimonopole with the conjugate spatial quantum number. It therefore places no requirement that the multiplicity of an individual fixed-$q=+1$ sector be even. For $\eta=\pm1,\pm2$, the associated Hermitian operators are most naturally organized as real representations built from the conjugate angular-momentum sectors; they need not be individual complex $C_6$ eigenoperators.

\begin{table*}[t]
\centering
\caption{\label{tab:monopole_decomposition}
Exact complex decomposition of the charge-one monopoles. $m_\eta$ is the complex dimension of the indicated $q=+1$ spin/$C_6$ sector and includes all $2S+1$ spin components. $m_{\Gamma,\eta}$ is its intersection with $T_1=T_2=1$, and $m_{{\rm non}\text{-}\Gamma,\eta}=m_\eta-m_{\Gamma,\eta}$. The last column gives the complex dimensions $d_+/d_-$ of the $\pm1$ eigenspaces of the dualized linear time-reversal coefficient map $\mathsf A_{\mathcal T}$ in the full indicated sector. Each complex $d_+$ ($d_-$) direction gives two real Hermitian time-reversal-even (time-reversal-odd) operators. For rows with $\eta=\pm1$ or $\eta=\pm2$, ``per sign'' means that the full ordered pair $(d_+,d_-)$ applies separately and identically to both signs of $\eta$; e.g. $(13,14)$ each means $(d_+,d_-)=(13,14)$ for both $\eta=+1$ and $\eta=-1$.
}
\setlength{\tabcolsep}{10pt}
\begin{tabular}{ccccccc}
\toprule
$S$ and $\SU(2)_s$ irrep & $\SU(6)_v$ irrep & $C_6$ sector
& $m_\eta$ & $m_{\Gamma,\eta}$ & $m_{{\rm non}\text{-}\Gamma,\eta}$ & $d_+/d_-$ \\
\midrule
$0\;(\mathbf 1_2)$ & $\mathbf{175}_6$ & $\eta=0$ & $32$ & $5$ & $27$ & $16/16$ \\
$0\;(\mathbf 1_2)$ & $\mathbf{175}_6$ & $\eta=3$ & $29$ & $2$ & $27$ & $13/16$ \\
$0\;(\mathbf 1_2)$ & $\mathbf{175}_6$ & $\eta=\pm1$ & $27$ per sign & $0$ & $27$ per sign & $13/14$ per sign
\\
$0\;(\mathbf 1_2)$ & $\mathbf{175}_6$ & $\eta=\pm2$ & $30$ per sign & $3$ per sign & $27$ per sign & $15/15$ per sign \\
\midrule
$1\;(\mathbf 3_2)$ & $\mathbf{189}_6$ & $\eta=0$ & $90$ & $3$ & $87$ & $45/45$ \\
$1\;(\mathbf 3_2)$ & $\mathbf{189}_6$ & $\eta=3$ & $99$ & $12$ & $87$ & $45/54$ \\
$1\;(\mathbf 3_2)$ & $\mathbf{189}_6$ & $\eta=\pm1$ & $99$ per sign & $12$ per sign & $87$ per sign & $45/54$ per sign \\
$1\;(\mathbf 3_2)$ & $\mathbf{189}_6$ & $\eta=\pm2$ & $90$ per sign & $3$ per sign & $87$ per sign & $45/45$ per sign \\
\midrule
$2\;(\mathbf 5_2)$ & $\mathbf{35}_6$ & $\eta=0$ & $30$ & $5$ & $25$ & $15/15$ \\
$2\;(\mathbf 5_2)$ & $\mathbf{35}_6$ & $\eta=3$ & $25$ & $0$ & $25$ & $10/15$ \\
$2\;(\mathbf 5_2)$ & $\mathbf{35}_6$ & $\eta=\pm1$ & $30$ per sign & $5$ per sign & $25$ per sign & $10/20$ per sign \\
$2\;(\mathbf 5_2)$ & $\mathbf{35}_6$ & $\eta=\pm2$ & $30$ per sign & $5$ per sign & $25$ per sign & $15/15$ per sign \\
\midrule
$3\;(\mathbf 7_2)$ & $\mathbf 1_6$ & $\eta=3$ & $7$ & $7$ & $0$ & $0/7$ \\
\bottomrule
\end{tabular}
\end{table*}

The parent dimensions are recovered exactly:
\begin{widetext}
\begin{align}
\dim(\mathbf 1_2,\mathbf{175}_6)
&=(5+27)_{\eta=0}+(2+27)_{\eta=3}
 +2(27)_{\eta=\pm1}+2(3+27)_{\eta=\pm2}=175,
\label{eq:S0_dimension_check}\\
\dim(\mathbf 3_2,\mathbf{189}_6)
&=(3+87)_{\eta=0}+(12+87)_{\eta=3}
 +2(12+87)_{\eta=\pm1}+2(3+87)_{\eta=\pm2}=567,
\label{eq:S1_dimension_check}\\
\dim(\mathbf 5_2,\mathbf{35}_6)
&=(5+25)_{\eta=0}+25_{\eta=3}
 +2(5+25)_{\eta=\pm1}+2(5+25)_{\eta=\pm2}=175,
\label{eq:S2_dimension_check}\\
\dim(\mathbf 7_2,\mathbf 1_6)&=7_{\eta=3}=7.
\label{eq:S3_dimension_check}
\end{align}
\end{widetext}
In each parenthesis, the first number is the translation-trivial
contribution and the second is the remaining nonzero-momentum
contribution. The factors of two count the distinct $+\eta$ and
$-\eta$ complex sectors.

\begin{table*}[t]
\centering
\caption{\label{tab:monopole_channels}
Translation-trivial charge-one sectors and representative microscopic channels within each $C_6$ sector. $m_\Gamma$ is the complex $q=+1$ dimension including spin components. $d_{\Gamma,+}$ and $d_{\Gamma,-}$ are the complex dimensions of the $\pm1$ eigenspaces of the dualized time-reversal coefficient map in the same sector. They correspond to $2d_{\Gamma,+}$ real Hermitian time-reversal-even and $2d_{\Gamma,-}$ real Hermitian time-reversal-odd operator directions.}
\setlength{\tabcolsep}{10pt}
\begin{tabular}{cccccp{6.8cm}}
\toprule
$S$ & $\SU(6)_v$ & $C_6$ sector & $m_\Gamma$ & $d_{\Gamma,+}/d_{\Gamma,-}$ & Representative channel(s) \\
\midrule
$0$ & $\mathbf{175}_6$ & $\eta=0$ & $5$ & $3/2$ & fully symmetric scalar ($6$ real allowed directions); T-odd scalar-chirality-type directions \\
$0$ & $\mathbf{175}_6$ & $\eta=3$ & $2$ & $0/2$ & staggered scalar-chirality-type directions \\
$0$ & $\mathbf{175}_6$ & $\eta=\pm2$ & $3$ per sign & $2/1$ per sign & $l=\pm2$ VBS-type components \\
\midrule
$1$ & $\mathbf{189}_6$ & $\eta=0$ & $3$ & $3/0$ & vector-spin-chirality-type channel \\
$1$ & $\mathbf{189}_6$ & $\eta=3$ & $12$ & $3/9$ & $l=3$ vector channels \\
$1$ & $\mathbf{189}_6$ & $\eta=\pm1$ & $12$ per sign & $3/9$ per sign & uniform $l=\pm1$ spin channels \\
$1$ & $\mathbf{189}_6$ & $\eta=\pm2$ & $3$ per sign & $3/0$ per sign & uniform $l=\pm2$ spin channels \\
\midrule
$2$ & $\mathbf{35}_6$ & $\eta=0$ & $5$ & $5/0$ & spin quadrupole \\
$2$ & $\mathbf{35}_6$ & $\eta=\pm1$ & $5$ per sign & $0/5$ per sign & $l=\pm1$ spin-quadrupole components \\
$2$ & $\mathbf{35}_6$ & $\eta=\pm2$ & $5$ per sign & $5/0$ per sign & $l=\pm2$ spin-quadrupole components \\
\midrule
$3$ & $\mathbf1_6$ & $\eta=3$ & $7$ & $0/7$ & spin-$3$ / octupolar channel \\
\bottomrule
\end{tabular}
\end{table*}

The stability-relevant sector is the $S=0$, $\eta=0$,
translation-trivial subspace, which has complex dimension five. The time-reversal coefficient map decomposes it as
\begin{equation}
5=3_+\oplus2_-.
\label{eq:central_T_split}
\end{equation}
One can choose a basis $V_a$, $a=1,2,3$, of the positive coefficient eigenspace
and define $\Phi_{{\rm triv},a}=V_a\cdot\Phi$. Physical time reversal
reverses the topological charge and acts as
\begin{equation}
\mathcal T\Phi_{{\rm triv},a}\mathcal T^{-1}
=\Phi_{{\rm triv},a}^{\dagger}.
\end{equation}
Because $\mathcal T$ is antiunitary, both
\begin{equation}
\Phi_{{\rm triv},a}+\Phi_{{\rm triv},a}^{\dagger},
\qquad
i\left(\Phi_{{\rm triv},a}-\Phi_{{\rm triv},a}^{\dagger}\right)
\end{equation}
are time-reversal even. Thus the three positive complex directions
supply six independent real Hermitian fully symmetric perturbations. The
two negative coefficient directions analogously supply four real
Hermitian time-reversal-odd operators. The most general symmetry-allowed
monopole perturbation is
\begin{equation}
\delta\mathcal L
=
\sum_{a=1}^{3}
\left(\lambda_a\Phi_{{\rm triv},a}+\lambda_a^{*}\Phi_{{\rm triv},a}^\dagger\right),
\qquad \lambda_a\in\mathbb C.
\label{eq:allowed_monopole_perturbation}
\end{equation}
Thus the maple-leaf DSL is not protected by symmetry exclusion of
charge-one monopoles; its stability depends on the scaling dimensions of
these six real Hermitian allowed perturbations. If relevant, their
proliferation may confine the gauge field without necessarily producing
conventional symmetry breaking, because the perturbations themselves
are fully symmetric.

The translation- and rotation-trivial sectors also contain one spin-$1$
multiplet, with three spin components, and one spin-$2$ multiplet, with
five spin components. They cannot appear as perturbations of an
$\SU(2)_s$-invariant Hamiltonian, but they are legitimate local
operators of the infrared theory and provide symmetry-resolved targets
for the excitation spectrum. Further spinful and
space-group-nontrivial monopoles occur in the remaining rows of
Tables~\ref{tab:monopole_decomposition} and
\ref{tab:monopole_channels}.

\subsection{Numerical detection}
\label{sec:numerical_detection}
The monopole classification gives concrete targets for numerical
studies of maple-leaf spin Hamiltonians. Because the monopole quantum
numbers translate directly into momentum, spin, and point-group sectors,
the field-theory classification can be tested in finite-size spectra of
microscopic Hamiltonians.  Starting from the six-valley mean-field
ansatz, one can construct the projected vacuum, neutral particle-hole
excitations, and flux-inserted monopole states, and compare their
overlaps with exact diagonalization eigenstates in the symmetry sectors
predicted by Tables~\ref{tab:monopole_decomposition} and~\ref{tab:monopole_channels}, following the strategy used in
Refs.~\cite{Wietek24,Budaraju-2023,Budaraju-2025}.

The six real Hermitian fully symmetric spin-singlet monopole perturbations associated with the three positive complex charge-one directions are the key operators for testing stability. In a deconfined DSL regime, flux-inserted variational states should exhibit the symmetry organization predicted by QED$_3$ and increasingly sharp separation from states in different emergent $Q_{\rm top}$ sectors with increasing system size. If monopoles proliferate, this flux-sector organization should break down, and the system may confine without producing Bragg peaks in either magnetic or valence-bond correlations.

Existing numerical studies motivate the search for a nonmagnetic maple-leaf regime, but they do not yet establish the $N_f=12$ QED$_3$ operator content, measure the charge-one monopole scaling dimension, or demonstrate emergent flux conservation. In particular, a variational preference for a gapped $\mathbb Z_2$ state represents a charge-two Higgs/spinon-pairing instability of the $\U(1)$ parent, rather than direct evidence for charge-one monopole proliferation. The present monopole classification should therefore be regarded as a set of symmetry-resolved tests for candidate microscopic Hamiltonians, not as an interpretation already confirmed by existing numerics.

Spinful and space-group-nontrivial monopoles provide further checks:
rather than focusing solely on static spin or valence-bond structure
factors, numerical studies should resolve the low-energy spectrum by
momentum, point group, and total spin, and ask whether the levels can be
assigned consistently to the monopole multiplets. The classification
suggests a practical search protocol: identify nonmagnetic regimes,
resolve the low-energy spectrum by symmetry, and compare the observed
levels with the QED$_3$ operator families predicted by the monopole and
bilinear quantum numbers.

A complementary diagnostic is the response to a time-reversal-breaking
scalar spin-chirality perturbation,
\(\mathbf S_i\cdot(\mathbf S_j\times\mathbf S_k)\).  In the DSL this
corresponds to the fermion mass \(\overline{\psi}\psi\), which gaps the
Dirac fermions and induces a Chern-Simons term for the emergent gauge
field, producing a chiral spin liquid with \(\SU(6)_1\) topological
order.  If, however, the symmetry-trivial monopoles have already
proliferated and the system is confined, the same perturbation need not
produce topological order without an intervening phase transition.

Finally, the momentum structure of monopoles in the present problem is
less rigid than in the triangular and kagome DSLs. In those cases, the
Dirac nodes are pinned to high-symmetry momenta, and the monopole
crystal momenta are correspondingly pinned to high-symmetry points in
the Brillouin zone~\cite{Song-2019,Song-2020}. For the restricted
nearest-neighbor maple-leaf ansatz of Eq.~\eqref{eq:H_mf}, the Dirac
nodes move continuously along the
$\overline{\vect{\Gamma}\mathrm{M}}$ lines according to
Eq.~\eqref{eq:exact_dirac_position}. This line confinement is not
required by the microscopic $p6$ space group: generic
symmetry-preserving longer-range hopping terms may displace the nodes
away from these lines while retaining a six-node orbit related by
$C_6$. Consequently, monopoles carrying nonzero crystal momentum need
not appear at fixed high-symmetry momenta. Their momenta depend on the
actual locations of the Dirac valleys in the microscopic band
structure. Numerically, low-lying singlet and spinful excitations
should therefore be sought at the momenta appropriate to the
Hamiltonian and finite cluster under consideration, rather than only
at the usual high-symmetry points. For the representative nearest-neighbor
point used in Fig.~\ref{fig:phase_diag}, the complete translation-diagonal
momentum decomposition is given in Appendix~\ref{app:monopole_representation}.

\section{Conclusion and Outlook}
\label{sec:conclusion}

We have classified the fundamental monopole operators of the maple-leaf-lattice $\U(1)$ Dirac spin liquid. The six-node ansatz realizes an $N_f=12$ QED$_3$ theory, but microscopic symmetries do not exclude charge-one monopole perturbations. In the stability-relevant spin-singlet, zero-momentum, $C_6$-trivial sector there are five complex charge-one directions. Time reversal splits them as $3_+\oplus2_-$; the three positive complex directions yield six independent real Hermitian monopole perturbations that are fully symmetric. The phase is therefore not guaranteed to be stable by symmetry. Its fate is controlled dynamically by the scaling dimensions of these allowed operators. Existing estimates lie close to marginality and do not determine whether the charge-one monopole is weakly relevant or weakly irrelevant at $N_f=12$.

This places the maple-leaf lattice in a different symmetry setting from the triangular and kagome examples. Here the minimum symmetry-allowed monopole charge is already one, while the number of Dirac fermion flavors is larger. The maple-leaf lattice therefore provides a concrete test of whether large-flavor dynamics can render symmetry-allowed charge-one monopoles irrelevant and stabilize a Dirac spin liquid.

The classification also gives concrete numerical consequences.  Besides
the fully symmetric spin-singlet monopole perturbations that control stability, we
find spinful and space-group-nontrivial monopole sectors that can be
searched for in symmetry-resolved spectra.  In analogy with recent work
on the triangular-lattice \(J_1\)-\(J_2\) antiferromagnet
~\cite{Wietek24,Budaraju-2023,Budaraju-2025}, exact diagonalization and
variational Monte Carlo studies should therefore test not only for the
absence of conventional order, but also for consistency with the
operator content of \(N_f=12\) QED$_3$: fermion bilinears,
flux-inserted monopole states, and their predicted momentum,
point-group, and spin quantum numbers.  In this way, the monopole
classification provides a spectral diagnostic for identifying, or
placing strong constraints on, a maple-leaf DSL regime.

Several directions remain open.  A more precise determination of the
charge-one monopole dimension at \(N_f=12\) would sharpen the stability
criterion, since current estimates lie close to marginality.  It would
also be valuable to search systematically for microscopic maple-leaf
Hamiltonians whose spectra and correlations are consistent with the
\(N_f=12\) DSL. This includes the nearest-neighbor Heisenberg model, as
well as models with further-neighbor exchange, ring exchange, or other
interactions that may stabilize the six-valley Dirac spectrum. In this
context, we have identified a maple-leaf model with a ring-exchange
term on the rhombi containing the dimer bonds for which preliminary
variational calculations point to a \(\U(1)\) Dirac spin-liquid ground
state; a detailed study will be presented elsewhere. More
broadly, our results position the maple-leaf lattice as a benchmark
case for the question of whether a two-dimensional \(\U(1)\) Dirac
spin liquid can be realized not through symmetry-forbidden monopoles,
but through the dynamics of large-flavor compact QED$_3$.

\section{Acknowledgements}
Y.Z. was
supported by a United States Department of Energy grant DE-SC0008739.
This work is supported by the Deutsche Forschungsgemeinschaft (DFG, German Research Foundation) through Project-ID 258499086 -- SFB 1170 and through the W\"urzburg-Dresden Cluster of Excellence on Complexity and Topology in Quantum Matter -- ctd.qmat Project-ID 390858490 -- EXC 2147. S.S. was supported by the U.S. National Science Foundation grant No. DMR 2245246 and by the Simons Collaboration on Ultra-Quantum Matter which is a grant from the Simons Foundation (651440, S.S.). This research was also supported in part by grant NSF PHY-2309135 to the Kavli Institute for Theoretical Physics and by the International Centre for Theoretical Sciences (ICTS) for participating in the Discussion Meeting - Fractionalized Quantum Matter (code: ICTS/DMFQM2025/07). Y.I. acknowledges support from the Abdus Salam International Centre for Theoretical Physics through the Associates Programme, from the Simons Foundation through Grant No.~284558FY19, from IIT Madras through the Institute of Eminence program for establishing QuCenDiEM (Project No. SP22231244CPETWOQCDHOC).

\appendix

\section{Details of the uniform U(1) Ansatz on the maple-leaf lattice}\label{app:psg}
Here, we reiterate the mean-field ansatz and PSG of the uniform $\U(1)$ spin liquid studied in~\cite{feuerpfeil_2026_higgs}:

We define the unit cell as consisting of six sites, indexed by $s=1, \dots, 6$. The position of any site is denoted by $(x,y,s)$, where the coordinates $(x,y)$ specify the unit cell location. The maple-leaf lattice belongs to the $p6$ wallpaper group, which is generated by two translations ($T_1$ and $T_2$) and a six-fold rotation ($C_6$) centered at the origin, as illustrated in \figref{fig:u1_qsl_maple}(a). The actions of these generators on the site coordinates $(x,y,s)$ are given by:
\begin{equation}
\left.\begin{aligned}
T_1:(x,y,s)&\rightarrow(x+1,y,s)\,,\\
T_2:(x,y,s)&\rightarrow(x,y+1,s)\,,\\
C_6:(x,y,s)&\rightarrow(x-y,x,C_6(s))\,.
\end{aligned}\right.
\end{equation}
Under the $C_6$ rotation, the sublattice indices are cyclically permuted according to the mapping $C_6(s) = (2, 3, 4, 5, 6, 1)$. The uniform $\mathrm{U}(1)$ {\it Ansatz} is defined by the following link and onsite fields:
\begin{equation}
\begin{split}
u^{}_{\text{hex}}&=t^{}_{h}\tau^z\,,\\
u^{}_{\text{dim}}&=t^{}_{d}\tau^z\,,\\
u^{}_{\text{tria}}&=t^{}_{t}\tau^z\,,\\
u^{}_{\text{onsite}}&=\lambda^{}_{z}\tau^z\,.\label{eq:maple_u1}
\end{split}
\end{equation}
The corresponding PSGs are given by:
\begin{equation}
\begin{split}
W^{}_{T_1}(x,y,s)&=e^{{i}\phi^{}_{T_1}\tau^z}\,,\\
W^{}_{T_2}(x,y,s)&=e^{{i}\phi^{}_{T_2}\tau^z}\,,\\
W^{}_{C_6}(x,y,s)&=e^{{i}\phi^{}_{C_6}\tau^z}\,,\\
W^{}_{\mathcal{T}}(x,y,s)&=i\tau^y e^{{i}\phi^{}_{\mathcal{T}}\tau^z}{i}\tau^x\,\,.\label{eq:maple_u1_PSG}
\end{split}
\end{equation}
Choosing $\phi_{\mathcal T}=\pi/2$ gives
\begin{equation}
i\tau^y e^{i(\pi/2)\tau^z}i\tau^x=\mathbbm 1,
\label{eq:PSG_T_cancellation}
\end{equation}
so the full microscopic time-reversal action on the spinons is the standard Kramers transformation
$f_{i\uparrow}\mapsto f_{i\downarrow}$ and
$f_{i\downarrow}\mapsto-f_{i\uparrow}$. It therefore reverses crystal momentum and, in the monopole sector, reverses $\U(1)_{\rm top}$ charge.

\subsection{Exact position of the six Dirac nodes}
\label{app:exact_nodes}

We derive here the exact node trajectory quoted in
Eq.~\eqref{eq:exact_dirac_position}.  Write $\vk=u\vb_1+w\vb_2$, so that
\begin{equation}
P=e^{2\pi i u},\qquad Q=e^{2\pi i w}.
\end{equation}
On the representative $\overline{\vect{\Gamma}\mathrm M}$ line one
has $w=0$ and hence $Q=1$.  Defining
\begin{equation}
    z=P=e^{2\pi i u},
\end{equation}
and subtracting the common diagonal term $\lambda_z$, the Bloch matrix
on this line is
\begin{equation}
\mathcal H(z)=
\begin{pmatrix}
0&t_h&t_tz&t_d&t_t&t_h\\
t_h&0&t_h&t_t&t_dz^{-1}&t_tz^{-1}\\
t_tz^{-1}&t_h&0&t_h&t_tz^{-1}&t_dz^{-1}\\
t_d&t_t&t_h&0&t_h&t_tz^{-1}\\
t_t&t_dz&t_tz&t_h&0&t_h\\
t_h&t_tz&t_dz&t_tz&t_h&0
\end{pmatrix}.
\label{eq:line_bloch_matrix}
\end{equation}

At the shifted eigenvalue
\begin{equation}
    \widetilde{\varepsilon}_D=-\frac{t_dt_t}{t_h},
\end{equation}
the determinant factorizes exactly as
\begin{equation}
\det\!\left[
\mathcal H(z)+\frac{t_dt_t}{t_h}\mathbbm 1_6
\right]
=
\frac{t_t^2-t_h^2}{t_h^6z^2}
\left[
t_dt_h^4z^2-\mathcal N z+t_dt_h^4
\right]^2,
\label{eq:exact_det_factorization}
\end{equation}
where
\begin{align}
\mathcal N
&=
t_d^3t_h^2-t_d^3t_t^2-t_dt_h^4
+3t_dt_h^2t_t^2+2t_h^5-2t_h^3t_t^2
\nonumber\\
&=
(t_h^2-t_t^2)(t_d^3+2t_h^3)
-t_dt_h^2(t_h^2-3t_t^2).
\label{eq:N_polynomial}
\end{align}
At a root $z=z_D$ of the quadratic factor in
Eq.~\eqref{eq:exact_det_factorization}, a direct symbolic evaluation
shows that every $5\times5$ minor of
\[
\mathcal M(z)
\equiv
\mathcal H(z)+\frac{t_dt_t}{t_h}\mathbbm 1_6
\]
is divisible by the same quadratic polynomial and therefore vanishes.
Hence $\operatorname{rank}\mathcal M(z_D)\leq4$.
On the other hand, the $4\times4$ minor formed from rows
$\{1,2,4,5\}$ and columns $\{1,3,4,6\}$ evaluates to
\[
-\frac{(t_d-t_h)^2(t_t^2-t_h^2)^2}{t_h^2},
\]
which is nonzero throughout the interior of the six-node regime.
Therefore, $\operatorname{rank}\mathcal M(z_D)=4$, and its kernel is
exactly two dimensional. Since $|z_D|=1$ and
$\mathcal H(z_D)$ is Hermitian, this proves that
\[
\widetilde{\varepsilon}_D=-\frac{t_dt_t}{t_h}
\]
is an exactly twofold-degenerate eigenvalue of the shifted Bloch
matrix, or equivalently that
$\varepsilon_D=\lambda_z-t_dt_t/t_h$ is an exactly
twofold-degenerate eigenvalue of the original $6\times6$ matrix,
away from the node-collision boundaries.

The node position is obtained by setting the quadratic
factor in Eq.~\eqref{eq:exact_det_factorization} to zero:
\begin{equation}
t_dt_h^4z^2-\mathcal N z+t_dt_h^4=0.
\end{equation}
Dividing by $t_dt_h^4z$ and using
$z+z^{-1}=2\cos(2\pi u)$ gives
Eq.~\eqref{eq:exact_dirac_position}.

Introducing the dimensionless ratios
\begin{equation}
    r=\frac{t_h}{t_d},
    \qquad
    s=\frac{t_t}{t_d},
\end{equation}
the node equation can equivalently be written as
\begin{equation}
\cos(2\pi u_D)
=
\frac{
r^2(1-r^2+2r^3)
-(1-r)^2(1+2r)s^2
}{
2r^4
}.
\label{eq:dimensionless_node}
\end{equation}
The two collision conditions follow from
\begin{align}
\cos(2\pi u_D)-1
&=
-\frac{(1-r)^2(1+2r)(s^2-r^2)}{2r^4},
\label{eq:gamma_collision}\\
\cos(2\pi u_D)+1
&=
\frac{r^2(1+r)(1-r+2r^2)}{2r^4}
\nonumber\\
&\quad-
\frac{(1-r)^2(1+2r)s^2}{2r^4}.
\label{eq:M_collision}
\end{align}
In the component containing $(r,s)=(-3/4,0)$, the conditions
$-1<\cos(2\pi u_D)<1$ reduce, for $-1<r<-1/2$, to
\begin{equation}
    |s|<|r|.
\end{equation}
This proves Eq.~\eqref{eq:six_node_region}.  The boundary
$|s|=|r|$ corresponds to $u_D=0$, where the six nodes collide at
$\vect{\Gamma}$.

\subsection{A $p6$-symmetric perturbation that moves the nodes off $\overline{\Gamma M}$}
\label{app:off_axis_hopping}
To demonstrate explicitly that the $\overline{\Gamma M}$ line pinning is not enforced by $p6$, consider a real longer-range hopping obtained by taking the $C_6$ orbit of one seed bond,
\begin{equation}
\delta H_{\rm LR}
=t_{\rm LR}\sum_{\bm R}\sum_{r=0}^{5}
\left(
f^\dagger_{\bm R,s_r}
 f_{\bm R+A^r(1,0),t_r}
+{\rm H.c.}\right),
\label{eq:off_axis_hopping}
\end{equation}
where
\begin{equation}
A=\begin{pmatrix}1&-1\\1&0\end{pmatrix},
\qquad s_r=1+r,\qquad t_r=4+r\pmod 6.
\end{equation}
The six cell displacements are
\begin{equation}
(1,0),(1,1),(0,1),(-1,0),(-1,-1),(0,-1).
\end{equation}
This perturbation is real, translation invariant, spin independent, and closed under $C_6$, and therefore preserves $p6$, time reversal, spin rotation, and the $\U(1)$ invariant gauge group. At the representative point $(t_h/t_d,t_t/t_d)=(-3/4,0)$, direct projection into the crossing subspace gives the leading displacement
\begin{equation}
(\delta u,\delta w)
=
(-0.0368733,\,1.055400)\,t_{\rm LR}+O(t_{\rm LR}^2).
\label{eq:off_axis_displacement}
\end{equation}
The nonzero transverse component proves explicitly that physical $p6$ symmetry does not pin the node to the $\overline{\Gamma M}$ line.

\section{Global symmetry of QED\texorpdfstring{$_\mbf{3}$}{3}}
\label{app:symmetry}
\subsection{Faithful IR symmetry: G\texorpdfstring{$_{\mathbf{\mathrm{IR}}}$}{IR}}
Let us review the symmetries of the continuum QED$_3$ theory that form the emergent IR symmetry group, $G_{\mathrm{IR}}$.
The theory $\mathcal{L}_{\mathrm{QED}_3}$ from Eq.~\eqref{eq:L_QED} has a continuous Lorentz group symmetry, $\SO(2,1)_L$, 
and the familiar discrete Lorentz symmetry actions of time-reversal $\mathcal{T}_{\mathrm{IR}}$, partial reflection $\mathcal{R}_{\mathrm{IR}}$, and charge conjugation $\mathcal{C}_{\mathrm{IR}}$.
The \textit{IR} subscript is a reminder that these bare actions in $G_{\mathrm{IR}}$ are not the same as the corresponding physical symmetry in $G_{\mathrm{UV}}$.
The discrete symmetries can be chosen to act on the Dirac fermions as
\begin{alignat}{1}
\mathcal{T}_{\mathrm{IR}}\;&:\; \psi(t,\vect{r})\rightarrow \gamma^1\psi^{\alpha}(-t,\vect{r})\,,\quad i\rightarrow -i\,,\label{eq:bareaction_dirac}\\
\mathcal{R}_{\mathrm{IR}}\;&:\;\psi(t,\vect{r})\rightarrow\gamma^2 \psi^{\alpha}(t,R\vect{r})\,,\\
    \mathcal{C}_{\mathrm{IR}}\;&:\; \psi(t,\vect{r})\rightarrow (\overline{\psi}^{\alpha}(t,\vect{r})\gamma^1)^T\,.
    \label{eq:bareaction_dirac_end}
\end{alignat}
Note, however, that the Dirac fermions are not local operators; we use them as a device to deduce the symmetry action on the monopoles, which are the true local operators of the theory.
Internally on the Dirac fermions, we have a $\mathrm{PSU}(12)_f$ flavor symmetry,
$\psi_{\alpha}\rightarrow U_{\alpha\beta}\psi_{\beta}$ for $U\in \mathrm{PSU}(12)_f$.
We have quotiented out the $\mathbb{Z}_{12}$ center of $\SU(12)_f$ as it can be combined with a gauge transformation.
In addition, we have a $\U(1)_{\mathrm{top}}$ symmetry associated with conservation of the gauge flux $\mathrm{d}a/2\pi$. This symmetry acts trivially on the Dirac fermions. 

To find the faithful global symmetry, we must examine the monopole operators $\Phi^{\dagger}\sim d_{i}^{\dagger}d_{j}^{\dagger}d_{k}^{\dagger}d_{l}^{\dagger}d_{m}^{\dagger}d_{n}^{\dagger}\mathcal{M}^{\dagger}$.
Since $\mathcal M$ is the flux vacuum and is taken to be flavor-symmetric,
while each zero-mode creation operator transforms in the fundamental
representation of $\SU(12)_f$, $\Phi$ transforms in the sixfold antisymmetric
tensor product of the $\SU(12)_f$ fundamental representation, which is the irreducible representation of $\SU(12)$ whose Young tableau has one column and six rows.
Furthermore, the fundamental monopoles $\Phi$ are Lorentz scalars.
Consequently, (in addition to the discrete and Lorentz symmetries mentioned previously) the true faithful internal symmetry group acting on the monopoles is
\begin{equation}
    G_{\mathrm{IR}}=\frac{\SU(12)_f\times \U(1)_{\mathrm{top}}}{\mathbb{Z}_{12}}\,,
\end{equation}
where $\mathbb{Z}_{12}$ is generated by the element $(e^{i \pi/6}\mathbb{I}_{12\times 12},-1)\in \SU(12)_f\times \U(1)_{\mathrm{top}}$.
The quotient arises because the generator $e^{i \pi/6}\mathbb{I}_{12\times12}$ of the center of $\SU(12)_f$ acts by $-1$ on a charge-one monopole in $\wedge^6\mathbf{12}$, exactly as does a $\pi$ rotation in $\U(1)_{\mathrm{top}}$.\footnote{In other words, any local operator's $\U(1)_{\mathrm{top}}$ charge $Q$ is related to its $N_f$-ality $n$ by $n(Q)=\frac{N_f\cdot Q}{2}\pmod{N_f}$.}

\subsection{Conventions for Dirac frames and symmetry transformations}
\label{app:dirac_frames}

These spinor matrices can be defined directly by projecting the
microscopic symmetry actions onto the two bands that cross at each
Dirac point. Let $V_v$ be a $6\times2$ orthonormal matrix whose
columns span the crossing subspace at $\vect K_v$. Furthermore, let

$\mathcal U_{C_6}(\vect k)$ and
$\mathcal U_{\mathcal T}(\vect k)$ denote the microscopic symmetry actions on the Bloch Hamiltonian,
\begin{align}
\mathcal U_{C_6}(\vect k)
H(\vect k)
\mathcal U_{C_6}^{\dagger}(\vect k)
&=
H(R_6\vect k),\\
\nonumber
\\
\mathcal U_{\mathcal T}(\vect k)
H^{*}(\vect k)
\mathcal U_{\mathcal T}^{\dagger}(\vect k)
&=
H(-\vect k).
\end{align}
The corresponding actions in the two-component low-energy subspaces
are
\begin{equation}
U_{C_6,v}
=
V_{v+1}^{\dagger}
\mathcal U_{C_6}(\vect K_v)
V_v,
\qquad
U_{\mathcal T,v}
=
V_{\overline{v}}^{\dagger}
\mathcal U_{\mathcal T}(\vect K_v)
V_{v}^{*}.
\label{eq:projected-Dirac-actions}
\end{equation}
These matrices are unitary because the microscopic symmetries map the
two-dimensional crossing subspace at one valley onto the corresponding
crossing subspace at the symmetry-related valley. 

The explicit matrix representatives $U_{C_6,v}$ and
$U_{\mathcal T,v}$ depend on the choice of the local two-component
Dirac frames. Under
\begin{equation}
V_v\longmapsto V_vW_v,
\qquad W_v\in\U(2).
\label{eq:Dirac-frame-change}
\end{equation}
they transform as
\begin{equation}
U_{C_6,v}
\longmapsto
W_{v+1}^{\dagger}U_{C_6,v}W_v,
\qquad
U_{\mathcal T,v}
\longmapsto
W_{\overline{v}}^{\dagger}U_{\mathcal T,v}W_v^{*}.
\label{eq:Dirac-frame-transformations}
\end{equation}
Their detailed form is therefore basis dependent, whereas the
resulting symmetry representation and all physical conclusions are
basis independent. In the common momentum frame used in the monopole
calculation, the gamma matrices are the same at every node and the
projected matrices are independent of $v$:
\begin{equation}
U_{C_6,v}=\mathbbm 1_2,
\qquad
U_{\mathcal T,v}=\rho_x.
\label{eq:Dirac_frame_final_matrices}
\end{equation}
Thus $C_6$ acts by the cyclic valley permutation after the Dirac-spinor
frame has been fixed, while physical time reversal reverses the valley
momentum and retains the residual $\rho_x$ action on the two-component
Dirac spinor.

\subsection{SU(12) tensor representations of the monopoles}
\label{app:monopole_su12}
Of the total \begin{equation}
    \begin{pmatrix}
    12\\6
\end{pmatrix}
=924
\end{equation} monopoles which transform in the sixfold fully antisymmetric tensor representation of $\SU(12)$, $\wedge^6\mathbf{12}=\mathbf{924}_{12}$, 
it is useful to fix a monopole basis that respects
the branching structure under 
$\SU(12)\rightarrow \SU(2)_s\times \SU(6)_v$, as the PSG factors through this branching.
In such a tensor product basis, we have the Young tableau decomposition
\begin{widetext}
\begin{alignat}{1}
    \mathbf{924}_{12}&=(\mathbf{7}_{2},\mathbf{1}_{6})\oplus (\mathbf{5}_2,\mathbf{35}_6)\oplus (\mathbf{3}_2,\mathbf{189}_6)\oplus (\mathbf{1}_2,\mathbf{175}_6)\,,\label{eq:SU12_branching}\\
    \ytableausetup{smalltableaux,aligntableaux=center}
\ydiagram{1,1,1,1,1,1}&=\left(\;\ydiagram{6}\,,\ydiagram{1,1,1,1,1,1}\;\right)\oplus\left(\;\ydiagram{5,1}\,,\ydiagram{2,1,1,1,1}\;\right)\oplus
\left(\;\ydiagram{4,2}\,,\ydiagram{2,2,1,1}\;\right)
\oplus\left(\;\ydiagram{3,3}\,,\ydiagram{2,2,2}\;\right)\,.
\end{alignat}

We now analyze each irreducible multiplet. In general, we can introduce tensors $V$ and $T$ transforming in the respective $\SU(2)_s/\SU(6)_v$ irreducible representations. Denoting the fermion zero-mode creation operators by $d^\dagger_{\sigma,i}$,
where $\sigma\in\{\uparrow,\downarrow\}$ and $i\in\{1,\ldots,6\}$, the
monopoles can be written schematically as
\begin{equation}
    \mathcal{A}((\sigma_1,i_1),\cdots,(\sigma_6,i_6))V_{\sigma_1\cdots\sigma_6}T_{i_1\cdots i_6}d_{\sigma_1,i_1}^{\dagger}d_{\sigma_2,i_2}^{\dagger}\cdots d_{\sigma_6,i_6}^{\dagger}\mathcal{M}_{1}^{\dagger}\,,
\end{equation}
\end{widetext}
where we have denoted $\mathcal{A}$ to completely antisymmetrize over its index arguments.
It is convenient to use the notation 
\begin{equation}
    {\Phi}_A^{\dagger}\,,
\label{eq:monopole_tensor_notation}
\end{equation}
where $A=[\alpha_1,\dots,\alpha_6]$ is an antisymmetric multi-index. 
The $\mathbf{924}_{12}$ representation is self-conjugate with an invariant
bilinear given by
\begin{equation}
   E_{AB}= E_{[\alpha_1,\dots,\alpha_6][\beta_1,\dots,\beta_6]}=\frac{1}{6!}\epsilon_{\alpha_1,\dots,\alpha_6,\beta_1,\dots,\beta_6}\,.
\label{eq:monopole_E_tensor}
\end{equation}
We remark that $E_{AB}=E_{BA}$, so the monopoles are in an orthogonal representation.

The simplest sector within this $924$-dimensional space is the $7$-dimensional $(\mathbf{7}_{2},\mathbf{1}_{6})$ manifold, which carries spin-$3$ and is an SU(6)$_v$ singlet. 
While these seven monopoles are valley singlets and therefore invariant (modulo U(1)$_{\mathrm{top}}$ actions) under the space group symmetries, they do carry spin.

The next sector is the $5\cdot35=175$-dimensional $(\mathbf{5}_2,\mathbf{35}_6)$ space, which has a spin-$2$ multiplet transforming in the adjoint of SU(6)$_v$.

The $(\mathbf{3}_2,\mathbf{189}_6)$ sector carries spin-$1$, with the valley transforming in the subspace of $\wedge^4 \mathbf{6}_6\otimes\wedge^2 \mathbf{6}_6$ minus the irreducible trace and adjoint components.
The $(\mathbf{1}_2,\mathbf{175}_6)$ sector is a spin singlet.
Therefore, as $\SU(2)_s$ symmetry is present, any symmetry-allowed monopole must come from this multiplet.

\subsection{Exterior-power construction and exact checks}
\label{app:monopole_representation}

We construct the charge-one monopole representation directly on the
exterior-power basis
\begin{equation}
\mathcal B=
\left\{I\subset\{1,\ldots,12\}:|I|=6\right\},
\qquad
|\mathcal B|=\binom{12}{6}=924.
\label{eq:wedge_basis}
\end{equation}
A one-particle unitary flavor transformation $U$ induces the matrix
$\wedge^6 U$ on this basis, including the fermionic sign required to
restore the canonical order of the occupied indices. We use this
construction directly for translations and rotation. Physical time reversal
changes the topological charge and is treated below after Hermitian
conjugation and dualization of the conjugate representation.

The physical charge-one rotation is the net operator
$\widehat C_6^{(q=1)}$ in Eq.~\eqref{eq:net_C6_monopole}. Its angular
momentum projector is
\begin{equation}
\Pi_\eta
=
\frac{1}{6}\sum_{n=0}^{5}
 e^{-i\pi\eta n/3}
\left(\widehat C_6^{(q=1)}\right)^n.
\label{eq:C6_monopole_projector}
\end{equation}
Let $\Pi_\Gamma$ be the joint spectral projector onto
$T_1=T_2=1$. Because $\Gamma$ is invariant under $C_6$,
$\Pi_\Gamma$ commutes with $\Pi_\eta$, although translations and
$C_6$ do not commute at a generic nonzero momentum.

To extract the total-spin content, let $d_m(\Pi)$ denote the dimension
of the image of a spatial projector $\Pi$ in the fixed
$S_z=m$ subspace. Since every spin-$S$ irrep contributes one state at
each $m=-S,\ldots,S$, the number of spin-$S$ multiplets is
\begin{equation}
n_S(\Pi)=d_{m=S}(\Pi)-d_{m=S+1}(\Pi),
\label{eq:spin_multiplet_extraction}
\end{equation}
and their full complex dimension is $(2S+1)n_S(\Pi)$. Thus,
\begin{align}
m_\eta(S)&=(2S+1)n_S(\Pi_\eta),\\
m_{\Gamma,\eta}(S)&=(2S+1)n_S(\Pi_\Gamma\Pi_\eta).
\end{align}
These formulas reproduce every entry of
Table~\ref{tab:monopole_decomposition} and the dimension identities in
Eqs.~\eqref{eq:S0_dimension_check}--\eqref{eq:S3_dimension_check}.

Let
\begin{equation}
P_{\mathcal T}
=
\begin{pmatrix}
0&0&0&1&0&0\\
0&0&0&0&1&0\\
0&0&0&0&0&1\\
1&0&0&0&0&0\\
0&1&0&0&0&0\\
0&0&1&0&0&0
\end{pmatrix}
\label{eq:opposite_valley_permutation}
\end{equation}
denote the opposite-valley permutation $(1\,4)(2\,5)(3\,6)$. Physical microscopic time reversal is antiunitary and reverses the topological charge. Because $\wedge^6\mathbf{12}$ is self-conjugate, the invariant tensor $E$ in Eq.~\eqref{eq:monopole_E_tensor} identifies the conjugate antimonopole representation with the same lower-index $924$-dimensional coefficient space. The linear coefficient-space map used in the numerical classification is therefore
\begin{equation}
\mathsf A_{\mathcal T}
=
e^{i\theta_{\mathcal T}}
\wedge^6\!\left(i\sigma^y\otimes P_{\mathcal T}\right)E,
\label{eq:T_coefficient_map}
\end{equation}
where the overall phase is fixed by the spin-$3$ reference-monopole convention discussed in Appendix~\ref{app:berry_phase}. In the convention used here, $\mathsf A_{\mathcal T}$ is real and obeys
\begin{equation}
\mathsf A_{\mathcal T}^2=\mathbbm1.
\label{eq:T_coefficient_square}
\end{equation}
In the row-vector convention of the notebook, we diagonalize
\begin{equation}
V\mathsf A_{\mathcal T}=+V,
\qquad
W\mathsf A_{\mathcal T}=-W.
\label{eq:T_coefficient_eigenvectors}
\end{equation}
For $\Phi_V=V\cdot\Phi$ and $\Phi_W=W\cdot\Phi$, physical time reversal gives
\begin{align}
\mathcal T\Phi_V\mathcal T^{-1}&=\Phi_V^\dagger,\\
\mathcal T\Phi_W\mathcal T^{-1}&=-\Phi_W^\dagger.
\label{eq:T_on_coefficient_monopoles}
\end{align}
Because $\mathcal T$ is antiunitary, the two real Hermitian operators
\begin{equation}
\Phi_V+\Phi_V^\dagger,
\qquad
i(\Phi_V-\Phi_V^\dagger)
\end{equation}
are both time-reversal even, whereas the analogous two combinations constructed from $\Phi_W$ are both time-reversal odd. Thus a complex $d_+$ coefficient eigenspace gives $2d_+$ real time-reversal-even Hermitian directions, and a complex $d_-$ eigenspace gives $2d_-$ real time-reversal-odd directions. Restricting Eq.~\eqref{eq:T_coefficient_map} to the $S=0$, $\Gamma$, $\eta=0$ sector gives the independently checked split $5=3_+\oplus2_-$ quoted in Eq.~\eqref{eq:central_T_split}.

The spatial representation independently satisfies
\begin{align}
\left(\widehat C_6^{(q=1)}\right)^6&=\mathbbm1,\\
\widehat C_6\widehat T_1\widehat C_6^{-1}&=\widehat T_1\widehat T_2,\\
\widehat C_6\widehat T_2\widehat C_6^{-1}&=\widehat T_1^{-1}.
\label{eq:monopole_group_algebra}
\end{align}

For completeness, we also diagonalize $T_1$ and $T_2$ directly at the
representative nearest-neighbor point
$(t_h/t_d,t_t/t_d)=(-3/4,0)$. There
\begin{equation}
\cos(2\pi u_D)=-\frac{13}{36},
\qquad
u_D=\frac{1}{2\pi}\arccos\!\left(-\frac{13}{36}\right).
\label{eq:representative_uD_momenta}
\end{equation}
The six valley momenta, in reduced reciprocal coordinates and in units
of $u_D$, are
\begin{equation}
(0,1),\;(-1,1),\;(-1,0),\;(0,-1),\;(1,-1),\;(1,0).
\label{eq:six_valley_reduced_coordinates}
\end{equation}
Summing six occupied zero-mode momenta gives $49$ distinct momenta,
organized into $\Gamma$ and eight generic six-point $C_6$ stars (orbits). Since
$-13/36$ is not among the rational cosine values allowed for a
rational multiple of $\pi$, $u_D$ is irrational; hence the listed
integer-coefficient momenta do not undergo accidental identifications
modulo reciprocal lattice vectors.

\begin{table*}[t]
\centering
\caption{\label{tab:monopole_momentum_stars}
Momentum decomposition of the monopoles at
$(t_h/t_d,t_t/t_d)=(-3/4,0)$. The second column gives one representative
$(u,w)$ of each $C_6$ momentum star, the third its orbit size, and
$n_S$ the number of spin-$S$ multiplets at each individual momentum in
the star. A row contributes $|\star|(2S+1)n_S$ complex states in spin
sector $S$.}
\setlength{\tabcolsep}{14pt}
\begin{tabular}{ccccccc}
\toprule
Momentum star & Representative $(u,w)$ & $|\star|$ & $n_{S=0}$ & $n_{S=1}$ & $n_{S=2}$ & $n_{S=3}$ \\
\midrule
$\Gamma$ & $(0,0)$ & $1$ & $13$ & $15$ & $5$ & $1$ \\
$\mathcal Q_1$ & $(0,-3u_D)$ & $6$ & $2$ & $2$ & $0$ & $0$ \\
$\mathcal Q_2$ & $(0,4u_D)$ & $6$ & $1$ & $0$ & $0$ & $0$ \\
$\mathcal Q_3$ & $(0,u_D)$ & $6$ & $8$ & $10$ & $2$ & $0$ \\
$\mathcal Q_4$ & $(0,-2u_D)$ & $6$ & $5$ & $4$ & $1$ & $0$ \\
$\mathcal Q_5$ & $(-3u_D,u_D)$ & $6$ & $2$ & $2$ & $0$ & $0$ \\
$\mathcal Q_6$ & $(-3u_D,2u_D)$ & $6$ & $2$ & $2$ & $0$ & $0$ \\
$\mathcal Q_7$ & $(4u_D,-2u_D)$ & $6$ & $1$ & $1$ & $0$ & $0$ \\
$\mathcal Q_8$ & $(u_D,u_D)$ & $6$ & $6$ & $8$ & $2$ & $0$ \\
\bottomrule
\end{tabular}
\end{table*}

The momentum table independently gives
\begin{align}
175&=13+6(2+1+8+5+2+2+1+6),\\
189&=15+6(2+0+10+4+2+2+1+8),\\
35&=5+6(0+0+2+1+0+0+0+2),\\
1&=1.
\label{eq:momentum_spin_checks}
\end{align}
Multiplying these spin-multiplet counts by $2S+1$ yields
$175,567,175,7$, and their sum is $924$. The $\Gamma$ row gives the
complex dimensions $13,45,25,7$, which agree with the sums of the
$m_{\Gamma,\eta}$ entries in Table~\ref{tab:monopole_decomposition}.
The original Mathematica notebook, an independent exterior-power
implementation, the machine-readable multiplicity tables, and
automated group-algebra checks are included in the accompanying source
archive.

\subsection{Discrete symmetry action on monopoles}
The elementary flux-reversing IR operations exchange monopoles and antimonopoles. In the conventions used below we choose
\begin{align}
\mathcal T_{\rm IR}&:\ \Phi_A\rightarrow E_{AB}\Phi_B^{\dagger},\\
\mathcal R_{\rm IR}&:\ \Phi_A\rightarrow E_{AB}\Phi_B^{\dagger},\\
\mathcal C_{\rm IR}&:\ \Phi_A\rightarrow\Phi_A^{\dagger}.
\end{align}
Here Hermitian conjugation converts the lower $\SU(12)_f$ index into the conjugate representation, and the invariant tensor $E$ returns it to the lower-index convention. The microscopic time-reversal operation also reverses $\U(1)_{\rm top}$ charge: it contains the physical Kramers spin rotation together with the opposite-valley and Dirac-spinor actions in Eq.~\eqref{eq:T-continuum}. Its action on charge-one coefficient vectors is therefore classified only after Hermitian conjugation and $E$-dualization, through the linear map $\mathsf A_{\mathcal T}$ of Eq.~\eqref{eq:T_coefficient_map}.

\section{The U(1)\texorpdfstring{$_{\mbf{top}}$}{top} action of G\texorpdfstring{$_{\mbf{UV}}$}{UV} symmetries}
\label{app:berry_phase}
\subsection{Berry phase of time-reversal}
Since the $\U(1)_{\mathrm{top}}$ phase is an inherently UV phenomenon, we trivialize the IR QED$_3$ by adding a quantum spin Hall mass term~\cite{Song-2020},
\begin{equation}
    \delta\mathcal{L}=m\overline{\psi}_i\sigma_z\psi_j\,, \quad m>0\,,
\end{equation}
which gaps out the spinons, leaving pure Maxwell gauge theory in the IR.
The zero-mode degeneracy of the monopoles is also lifted, leaving a unique light monopole $\Phi_{\mathrm{light}}$ in which all six valley zero-modes are occupied with spin $\downarrow$; this assignment is fixed by symmetry.
The mass term breaks the flavor symmetry down to $\SU(12)\rightarrow \SO(2)\times \dots$, where $\SO(2)$ is the conserved spin along the $z$-axis.
Probing the theory (or weakly gauging the $\SO(2)$) with an $\SO(2)$ gauge field $\mathcal{A}^{\SO(2)}$, the massive fermions will generate a topological term
\begin{equation}
    \mathcal{L}_{\mathrm{top}}=\frac{3}{2\pi} \mathcal{A}^{\SO(2)}\wedge \mathrm{d}a\,,\label{eq:topological_mass_response}
\end{equation}
which means the monopole carries charge $3$ under $\SO(2)$ (and is a singlet under the rest of the unbroken flavor symmetry).
Now in this regime, we can calculate the Berry phase of light monopole in this theory, from which by $\SU(12)$ rotations, we can find the transformations of the other monopoles.

The added mass $\delta \mathcal{L}$ leads to a quantum spin Hall insulating state. A single quantum spin Hall insulator is a $\mathbb{Z}_2$ topological insulator~\cite{Kane-2005} protected by Kramers time reversal. In the convention used here, a single copy takes the corresponding light monopole to minus its Hermitian conjugate. Because the maple-leaf theory contains three such copies, the reference spin-$3$, valley-singlet monopole obeys
\begin{equation}
\mathcal T\Phi_{\rm light}\mathcal T^{-1}=-\Phi_{\rm light}^{\dagger}.
\label{eq:T_light_monopole}
\end{equation}
The same sign is consistent with the topological response in Eq.~\eqref{eq:topological_mass_response}, for which a charge-one monopole carries spin $3$. This fixes the overall reference phase used in the coefficient map $\mathsf A_{\mathcal T}$ of Eq.~\eqref{eq:T_coefficient_map}. The physically important statement is that microscopic time reversal reverses $\U(1)_{\rm top}$ charge; the $924\times924$ matrix diagonalized in the classification is the dualized linear map on charge-one coefficient vectors, not a physical time-reversal endomorphism of the $q=+1$ operator space.

\subsection{Details of the Wannier center decomposition}
We determine the space-group Berry phases by decomposing the gapped
spinon bands, after adding a small quantum spin Hall mass, into
Wannier-center representations~\cite{Song-2020}.
Concretely, this is done by computing the eigenvalues of rotations at high-symmetry points in momentum space.
The $\U(1)_{\mathrm{top}}$ phase can then be thought of as arising from the Aharonov-Bohm phase acquired by a monopole moving in the insulating charge background of the Fermi sea.
A Wannier insulator of charge $q_r$ centered at a rotation center $r$ means that for a $C_n$ rotation around $r$, a monopole will pick up a phase $e^{iq_r2\pi/n}$. 
The Berry phase associated with translation symmetry can be determined by decomposing translation into compositions of rotations.
As the spinon bands are gapless, we add a small quantum spin Hall mass perturbation to gap out the bands.
Such a mass will not affect the monopole quantum numbers under the spatial symmetries and can be implemented by adding a directed imaginary hopping along the $t_t$ bonds, with opposite signs for the two spin species.

We compare the resulting spinon band insulator to a basis of Wannier insulators on the maple-leaf lattice. To begin, we observe the maple-leaf lattice space group is $p6$, generated by two translations $T_{1,2}$ and $C_6$ rotation around a hexagonal plaquette.
The relevant high-symmetry momenta (in reciprocal lattice units) are $\vect{\Gamma}=(0,0)$, $\vect{K}=(\frac{1}{3},\frac{1}{3})$, $\vect{K}'=C_6(\vect{K})$ and $\vect{M}=(\frac{1}{2},0)$.

We denote by $\Gamma^{\rm PSG}$ the formal band representation of the occupied spinon bands of the PSG mean-field Hamiltonian, after adding the small quantum spin Hall mass used to gap the Dirac nodes. Equivalently, $\Gamma^{\rm PSG}$ is the Wannier-center charge background of the filled spinon sea, expressed as an integer linear combination of elementary atomic band representations. This charge background is the object whose rotation-center charges determine the $\mathrm{U}(1)_{\rm top}$ Berry phases of monopoles.

The possible $p6$ Wannier centers are the usual ones centered on the $C_6$ center (hexagon), $C_3$ centers (triangular plaquettes defined by the $t_t$ bonds), $C_2$ centers (the $t_d$ bond centers), and the physical sites, which form a $C_6$ orbit.
Accordingly, we analyze the Wannier insulators centered on the hexagonal plaquettes $(\Gamma^h)$, triangular plaquettes $(\Gamma^t)$, bond centers $(\Gamma^{e})$, and sites $(\Gamma^v)$. 
The representation of $\Gamma^v$ is immediate, since it is the original tight-binding basis. For the other insulators, we represent them by taking equal-amplitude superpositions of spinons at the boundary sites of the relevant plaquette/triangle/bond.

We consider the eigenvalues under the original rotation $C_{6,h}$ (in addition to $C_{3h}$ and $C_{2h}$).
There is also rotation around a $C_3$ triangular plaquette, $C_{3t}$, and rotation around a $C_2$ bond, $C_{2e}$.
However, we do not consider these operations as they are related to $C_{6,h}$ by translation and do not provide additional information.
For more complicated PSGs or enlarged unit cells, simply matching eigenvalues may not be sufficient; one should in principle compare the full space-group or PSG representation.

We derive how to extract these eigenvalues in the general case. 
Given eigenvectors of the second quantized Hamiltonian $h$,
\begin{equation}
    h(k)\xi_n(k)=E_n(k)\xi_n(k)\,,
\end{equation}
we can collect the occupied eigenvectors in $\Xi(k)=(\xi_1(k),\dots,\xi_{N_{\rm occ}}(k))$.
For a symmetry $g$ that acts on the spinon Bloch operators as
\begin{equation}
g f_k g^{-1}=D_g(k)f_{gk}\,,
\end{equation}
the matrix $D_g(k)$ is the single-particle symmetry matrix in the spinon Bloch basis. It includes the sublattice permutation, the associated Bloch phase factors, and the projective gauge transformation required by the PSG action of $g$. The induced action of $g$ on the annihilation operators of the occupied bands is then
\begin{equation}
\begin{split}
    \Xi^\dagger(k)f_k&\rightarrow
    \Xi^\dagger(k)D_g(k)f_{gk}\\
    &\qquad=
    \Xi^\dagger(k)D_g(k)\Xi(gk)
    \left[\Xi^\dagger(gk)f_{gk}\right]\,.
\end{split}
\end{equation}
Equivalently, the occupied-band wavefunctions transform with the matrix
\begin{equation}
\Xi^\dagger(gk)D_g(k)^\dagger\Xi(k)\,.
\end{equation}
Diagonalizing this matrix at symmetry-invariant momenta, or within the appropriate little group of $k$, yields the symmetry eigenvalues of the occupied spinon bands.

Let
\begin{equation}
\Omega=e^{2\pi i/6},\qquad \omega=e^{2\pi i/3}=\Omega^2 .
\end{equation}
The hexagon-centered Wannier representations are denoted by $\Gamma^h_\eta$, where $\eta\in\mathbb Z_6$ labels the intrinsic $C_{6h}$ angular momentum, so that a Wannier orbital at the hexagon center transforms with eigenvalue $\Omega^\eta$ under $C_{6h}$. Similarly, $\Gamma^t_\eta$, with $\eta\in\mathbb Z_3$, denotes a Wannier representation centered on the triangular plaquettes, with intrinsic $C_{3t}$ angular momentum $\omega^\eta$. Since, $C_{6h}$ exchanges the two triangular plaquette centers in a unit cell, $\Gamma^t_\eta$ is a two-band representation. At the $\Gamma$ point, for example, the corresponding $C_{6h}$ action takes the form
\begin{equation}
\label{eq:C6h_triangular_wannier_action}
\begin{pmatrix}
0 & 1\\
\omega^\eta e^{i\mathbf{k}\cdot\mathbf{a}_1} & 0
\end{pmatrix}.
\end{equation}
This gives the $C_{6h}$ eigenvalues $\{\Omega^\eta,\Omega^{\eta+3}\}$. The bond-centered representations are denoted by $\Gamma^e_l$, where $l\in\mathbb Z_2$ labels the intrinsic $C_{2e}$ angular momentum $(-1)^l$. Finally, $\Gamma^v$ denotes the atomic representation obtained from Wannier orbitals on the six physical sites, which form a single $C_6$ orbit. For these bond-centered representations, we write $\lambda=(-1)^l$. The rotation eigenvalues of these elementary Wannier-center representations are listed in Table~\ref{tab:wannier_table}.

\begin{widetext}

\begin{table}[h]
	\centering
	\begin{tabular}{ |c||c|c|c|c||c|} 
		\hline
		$ $ & $ \Gamma^{h}_{\eta} $ & $ \Gamma^{t}_{\eta} $ & $ \Gamma^{e}_{l} $ &
        $\Gamma^{v}$&
        \text{HSM}\\[5pt] \hline\hline
		$ C_{6h} $ & $ \{\Omega^{\eta}\} $ & $ \{\Omega^{\eta},\Omega^{\eta+3}\} $ & $ \{\Omega^{l},\Omega^{l+2},\Omega^{l+4}\}$ & $\{1,\Omega,\Omega^2,\Omega^3,\Omega^4,\Omega^5\}$	& $\vect{\Gamma}$
        \\[5pt] \hline
		$ C_{3h} $& $ \{\omega^{\eta}\} $ & $ \{\omega^{\eta+1},\omega^{\eta+2}\}$ & $ \{1,\omega,\omega^2\}$ & $\oplus^2\{1,\omega,\omega^2\}$	& $\vect{K},\vect{K}'$\\[5pt] \hline
		$ C_{2h} $ & $ (-1)^{\eta}  $ & $ \{\pm1\}$ & $ \{\lambda,-\lambda,-\lambda\}$& $\oplus^3\{\pm1\}$& $\vect{M}$	\\[5pt] \hline
		\hline
	\end{tabular}
	\caption{\label{tab:wannier_table} Space-group rotation eigenvalues of the elementary Wannier-center representations on the maple-leaf lattice. The columns correspond to Wannier orbitals centered on the hexagon center $\Gamma^h_\eta$, triangular plaquette center $\Gamma^t_\eta$, bond center $\Gamma^e_l$, and physical site orbit $\Gamma^v$.  The entries give the rotation eigenvalues at the high-symmetry momenta $\boldsymbol{\Gamma}$, $\mathbf K,\mathbf K'$, and $\mathbf M$, and form the basis used to decompose the gapped spinon bands and extract the monopole Berry phases.}
\end{table}
\end{widetext}
Diagonalization of the spinon Hamiltonian leads to the representation of the spinon bands, shown in Table~\ref{tab:spinon_eig_table}. 
As expected, we can see that from the product of the eigenvalues $\xi_{C_{6h}}\xi_{C_{3h}}\xi_{C_{2h}}\ne1$, the spinon bands for a single sector are topological and cannot be decomposed in terms of Wannier insulators. 
Instead, the occupied bands exhibit fragile topology~\cite{Po-2018}; even though they are topologically trivial in the strictest sense, they can be deformed into a Wannier limit only after being combined with another atomic insulator.
\begin{table}[h]
	\centering
	\begin{tabular}{ |c||c||c|} 
		\hline
		$ $ &$\Gamma^{\mathrm{PSG}}$&
        \text{HSM}\\[5pt] \hline\hline
		$ C_{6h} $ & $ \oplus^2\{1,\Omega,\Omega^{-1}\} $ 	& $\vect{\Gamma}$
        \\[5pt] \hline
		$ C_{3h} $& $ \oplus^2\{1,\omega,\omega^2\} $ & $\vect{K},\vect{K}'$\\[5pt] \hline
		$ C_{2h} $ & $ \oplus^2\{-1,1,1\}  $ & $\vect{M}$	\\[5pt] \hline
		\hline
	\end{tabular}
	\caption{\label{tab:spinon_eig_table}Space-group rotation eigenvalues of the occupied spinon bands for the $N_f=12$ maple-leaf DSL at high-symmetry momenta. The overall multiplicity $\oplus^2$ arises from spin degeneracy. Comparing these eigenvalues with the Wannier-center representations in Table~\ref{tab:wannier_table} yields the effective Wannier decomposition used to determine the $\mathrm{U}(1)_{\rm top}$ Berry phases of lattice symmetries.}
\end{table}

We now determine $\Gamma^{\rm PSG}$ by matching the rotation eigenvalues of the occupied spinon bands in Table~\ref{tab:spinon_eig_table} to the elementary Wannier-center representations in Table~\ref{tab:wannier_table}. This matching is a linear problem in the integer coefficients of the atomic representations. One convenient solution is
\begin{equation}
\Gamma^{\rm PSG}
=
2\Gamma^h_0+\Gamma^h_1-\Gamma^h_3+\Gamma^h_5-\Gamma^e_0+\Gamma^v .
\label{eqn:decomposition}
\end{equation}
The equality here is meant at the level of the space-group rotation eigenvalue data.  Indeed, the right-hand side gives the $C_{6h}$
eigenvalues
\[
\oplus^2\{1,\Omega,\Omega^{-1}\}
\]
at $\Gamma$, the $C_{3h}$ eigenvalues
\[
\oplus^2\{1,\omega,\omega^2\}
\]
at $K$ and $K'$, and the $C_{2h}$ eigenvalues
\[
\oplus^2\{-1,1,1\}
\]
at $M$, in agreement with Table~\ref{tab:spinon_eig_table}.

The decomposition in Eq.~\eqref{eqn:decomposition} assigns a net effective gauge charge $q_{C_{6h}}=3$ to the hexagon-centered $C_6$ rotation center, modulo six. Therefore, a $2\pi$ monopole acquires the Berry phase
\begin{equation}
\exp\!\left(\frac{2\pi i q_{C_{6h}}}{6}\right)
=
\exp(i\pi)
=
-1
\end{equation}
under $C_{6h}$. The translation Berry phases are trivial. For example, using $T_1=C_{3t}C_{6h}^4$, the $C_{6h}^4$ contribution is $\exp(2\pi i\,q_{C_{6h}}\,4/6)=1$, and there is no net $C_{3t}$-centered charge contribution. Similarly, $T_2=C_{2h}C_{2e}$ gives a trivial phase because the two $C_2$ rotation-center contributions cancel
modulo two.

The decomposition in Eq.~\eqref{eqn:decomposition} is not unique. The reason is that the rotation eigenvalues in Tables~\ref{tab:wannier_table} and \ref{tab:spinon_eig_table} do not distinguish certain combinations of atomic representations. In particular, the following differences have identical rotation eigenvalue data at the high-symmetry momenta:
\begin{equation}
\sum_{\eta=0}^{2}\Gamma^t_\eta-\sum_{\eta=0}^{5}\Gamma^h_\eta,
\qquad
\Gamma^e_0+\Gamma^e_1-\sum_{\eta=0}^{5}\Gamma^h_\eta,
\qquad
\Gamma^v-\sum_{\eta=0}^{5}\Gamma^h_\eta .
\end{equation}
Thus, one may add
\begin{equation}
\begin{aligned}
& a\left(
\sum_{\eta=0}^{2}\Gamma^t_\eta-\sum_{\eta=0}^{5}\Gamma^h_\eta
\right)
+b\left(
\Gamma^e_0+\Gamma^e_1-\sum_{\eta=0}^{5}\Gamma^h_\eta
\right) \\
&\hspace{2.5em}
+c\left(
\Gamma^v-\sum_{\eta=0}^{5}\Gamma^h_\eta
\right) .
\end{aligned}
\label{eqn:arb_sum}
\end{equation}
where $a,b,c$ are arbitrary integers. Adding such a combination does not
change the rotation eigenvalues used in the matching. This non-uniqueness does not affect the monopole Berry phases, because these phases depend only on the rotation-center charges modulo the corresponding rotation order:
$q_{C_{6h}}$ modulo $6$, $q_{C_{3t}}$ modulo $3$, and $q_{C_{2e}}$ modulo $2$. Each of the three differences in Eq.~\eqref{eqn:arb_sum} changes these charges only by multiples of the corresponding rotation order.
The three differences displayed in Eq.~\eqref{eqn:arb_sum} are in fact a complete integral basis of the ambiguity. Writing the six $h$-center coefficients, three $t$-center coefficients, two $e$-center coefficients, and the $v$ coefficient as a twelve-component integer vector, the matching matrix for the six $\Gamma$-point $C_6$ multiplicities, three $K$-point $C_3$ multiplicities, and two $M$-point $C_2$ multiplicities has rank nine. Exact integer row reduction leaves three free variables and gives the three generators in Eq.~\eqref{eqn:arb_sum} as a saturated basis of the kernel. Their changes in the effective charges at the $(C_6,C_3,C_2)$ centers are respectively
\begin{equation}
(-6,3,0),\qquad(-6,0,2),\qquad(-6,0,0),
\label{eq:wannier_kernel_charge_changes}
\end{equation}
which vanish in $\mathbb Z_6\oplus\mathbb Z_3\oplus\mathbb Z_2$.
The translation phases are independent of the ambiguity as well. For $T_1=C_{3t}C_{6h}^4$, the change of the phase exponent is
\begin{equation}
\delta\nu_1=\frac{\delta Q_t}{3}+\frac{4\,\delta Q_h}{6},
\end{equation}
and for $T_2=C_{2h}C_{2e}$ it is
\begin{equation}
\delta\nu_2=\frac{\delta Q_h}{2}+\frac{\delta Q_e}{2}.
\end{equation}
For each of the three kernel generators these quantities are integers. Hence every rotation and translation Berry phase, including all relative signs, is independent of the arbitrary integers $a,b,c$.

\section{The unstable mean-field DSL}
\label{app:dsl_2}
\begin{table}[b]
	\centering
	\begin{tabular}{ |c||c||c|} 
		\hline
		$ $ &$\Gamma^{\mathrm{PSG}}$&
        \text{HSM}\\[5pt] \hline\hline
		$ C_{6h} $ & $ \{-1,\Omega^2,\Omega^{5}\}\oplus\{-1,\Omega^1,\Omega^{4
        }\} $ 	& $\vect{\Gamma}$
        \\[5pt] \hline
		$ C_{3h} $& $ \oplus^2\{1,\omega,\omega^2\} $ & $\vect{K},\vect{K}'$\\[5pt] \hline
		$ C_{2h} $ & $ \oplus^2\{-1,1,1\}  $ & $\vect{M}$	\\[5pt] \hline
		\hline
	\end{tabular}
	\caption{\label{tab:dsl2_eig_table} Space-group rotation eigenvalues of the occupied spinon bands for the distinct $N_f=4$ mean-field DSL appearing at $t_h=t_d$ and $t_t=0$. This state has two Dirac nodes at $\boldsymbol{\Gamma}$ and is discussed only for comparison, since it is unstable already at the mean-field level to infinitesimal changes of the hopping parameters.}
\end{table}

At $t_h=t_d=1$ and $t_t=0$, \figref{fig:phase_diag} shows a DSL that is a distinct phase from the DSL considered in the main text. In particular, this DSL has two low-energy Dirac nodes at $\vect{\Gamma}$, so its IR is described by $N_f=4$ QED$_3$.
For completeness, we perform a similar analysis of its monopoles and instabilities. The instability at the mean-field level refers to the absence of symmetry
protection for the Dirac nodes at $\Gamma$: generic allowed changes of the
hopping parameters move the system away from this fine-tuned band touching
and open a gap.

The lattice symmetries act on the Dirac fermions as
\begin{align}
    T_{1,2}&\;:\; \psi\rightarrow \psi\,,\\
    C_6&\;:\; \psi\rightarrow i\mu_ze^{-i\pi\rho_z/6}\psi\,,\\
    \mathcal{T}&\;:\; \psi\rightarrow i\sigma_y\mu_y\rho_y\psi\,.
\end{align}
Here, $\rho$ acts on the two-component Dirac spinor structure, while $\mu$ acts on the two-valley/node space of the $N_f=4$ DSL.

The filled bands  in the presence of a quantum spin Hall mass can be described by a conventional Wannier decomposition (see Table~\ref{tab:dsl2_eig_table}),
\begin{equation}
    \Gamma^{\mathrm{PSG}}=\Gamma^h_2+\Gamma^h_3+
    \Gamma^h_4+
    \Gamma_1^e\,,
\end{equation}
which leads to the same Berry phase contributions as the $N_f=12$ DSL.

With $N_f=4$, there are 
\begin{equation}
    {{4}\choose{2}}=6
\end{equation}
fundamental monopoles, which are strongly relevant and transform under the global symmetry
\begin{equation}
    \frac{\SU(4)_f\times \U(1)_{\mathrm{top}}}{\mathbb{Z}_4}\,.
\end{equation}
In particular, the monopoles organize as an $\SO(6)$ vector under $\SU(4)/\mathbb{Z}_2\simeq \SO(6)$.
The monopoles, with $\SO(6)$ vector indices, transform as in Table~\ref{tab:dsl2_monopoles}.
\begin{table}[h]
	\centering
	\begin{tabular}{ |c||c|c|} 
		\hline
		 &$C_6$&
        $\mathcal{T}$\\[5pt] \hline\hline
		$ \Phi_1$ & $\Phi_1$	& $\Phi_1^{\dagger}$
        \\[5pt] \hline
		$ \Phi_2$ & $\Phi_2$	& $\Phi_2^{\dagger}$
        \\[5pt] \hline
		$ \Phi_3$ & $-\Phi_3$	& $\Phi_3^{\dagger}$
        \\[5pt] \hline
        $ \Phi_4$ & $-\Phi_4$	& $-\Phi_4^{\dagger}$
        \\[5pt] \hline
        $ \Phi_5$ & $-\Phi_5$	& $-\Phi_5^{\dagger}$
        \\[5pt] \hline
        $ \Phi_6$ & $-\Phi_6$	& $-\Phi_6^{\dagger}$
        \\[5pt] \hline
		\hline
	\end{tabular}
	\caption{\label{tab:dsl2_monopoles} Transformation properties of the charge-one monopole operators of the unstable $N_f=4$ mean-field DSL under $C_6$ and time reversal $\mathcal T$. All monopole operators are trivial under translations. The table is included for comparison with the $N_f=12$ maple-leaf DSL analyzed in the main text.}
\end{table}

Thus, two monopoles, \(\Phi_{1,2}\), are symmetry trivial and their real and imaginary parts form a four dimensional allowed subspace of monopole operators. Together with the symmetry-allowed valley Hall mass, namely a Dirac mass with opposite signs in the two valleys that is already present at mean-field level, this shows that the $N_f=4$ DSL is not a stable phase. If the monopoles proliferate, the resulting confined phase is expected to
spontaneously break the residual (anomalous) \(\SO(4)\) IR symmetry, as in related examples~\cite{Zhang-2026b}.

\section*{Data availability}
Machine-readable tables of all $C_6$, momentum-star, and
time-reversal coefficient multiplicities reported in this work are
openly available in the Zenodo repository at
\url{https://doi.org/10.5281/zenodo.22067874}.

\section*{Code availability}
The Mathematica notebook used for the $924$-dimensional monopole
classification, together with two mutually independent Python
verification programs that reconstruct the exterior-power
representation and perform automated dimension and group-algebra
checks, is openly available in the Zenodo repository at
\url{https://doi.org/10.5281/zenodo.22067874}.

\clearpage

\bibliography{refs.bib}

\end{document}